\documentclass[
superscriptaddress,
nofootinbib, twocolumn,
amsmath,amssymb,aps,
floatfix, eprint ]{revtex4-2}

\usepackage{graphicx}
\usepackage{dcolumn}
\usepackage{bm}
\usepackage{amsmath,amssymb,mathrsfs,amsfonts,slashed}
\usepackage{hyperref,xfrac}
\usepackage{url}
\usepackage{xcolor}
\usepackage[normalem]{ulem}
\newcommand{\p}{\bm{p}}

\begin{document}

\title{T-odd Parton Distribution Functions and Azimuthal Anisotropy at High Transverse Momentum in $p$-$p$ and $p$-$A$ Collisions }

\author{Ismail Soudi}
\email{ismail.soudi@wayne.edu}
\affiliation{ Department of Physics and Astronomy, Wayne State University, Detroit, MI 48201. }
\affiliation{University of Jyväskylä, Department of Physics, P.O. Box 35, FI-40014 University of Jyväskylä, Finland.}
\affiliation{Helsinki Institute of Physics, P.O. Box 64, FI-00014 University of Helsinki, Finland.}

\author{Abhijit Majumder}%
\email{majumder@wayne.edu}
\affiliation{ Department of Physics and Astronomy, Wayne State University, Detroit, MI 48201. }
\date{\today}
\begin{abstract}
Various azimuthal anisotropies ($v_1, v_2, v_3, v_4$), at high transverse momentum (high-$p_T$), are shown to arise from the asymmetric scattering of transverse polarized quarks and gluons, arising from unpolarized nucleons (the Boer-Mulders' effect) and resulting in unpolarized hadrons (the Collins effect). Combined with the asymmetric scattering of partons from polarization independent but transverse momentum dependent (TMD) distributions, we obtain a possible mechanism to understand the azimuthal anisotropy of hadrons at large transverse momentum observed in $p$-$p$ collisions. Constraining the ratio of polarization dependent TMD distributions to polarization independent distributions by comparing with the data from $p$-$p$ collisions, we find that scaling the acquired transverse momentum of the initial state partons from the proton, due to prescattering before the hard interaction, with the length of the nucleus ($k_\perp^2 \propto A^{1/3}$), straightforwardly yields the azimuthal anisotropy at high $p_T$ in $p$-$A$ collisions. 
\end{abstract}

\maketitle
\section{Introduction}
One of the foremost signals of the formation of a quark-gluon plasma (QGP) during a heavy ion collision is the elliptic flow of the bulk medium~\cite{STAR:2004jwm,PHENIX:2003qra,ALICE:2010suc,CMS:2012zex,ATLAS:2012at}, in semi-central collisions. 
\footnote{Centrality of the collision indicates the degree of transverse spatial overlap of the two incoming nuclei (with $0\%$ indicating most central or maximal overlap, and $100\%$ indicating most peripheral or a grazing collision).} 
Due to the geometry of the collision, the overlap region of the two nuclei is not circular, and contains azimuthal anisotropies in density. 
Pressure gradients created by these spatial inhomogeneities lead to azimuthal anisotropies in the momentum distribution~\cite{Kolb:2000sd}, quantified by the Fourier coefficients of the differential hadronic yield, 
\begin{align}
\frac{dN}{d\phi} = \frac{dN}{d\phi} \left(1+ \sum_n 2 v_n \cos[n(\phi-\Psi_n)]\right) \;.
\end{align}
In the above equation, $\phi$ is the azimuthal angle and $\Psi_n$ is the $n^{\rm th}$ order event plane angle, and $dN$ is the differential yield of hadrons, in some range of transverse momentum and rapidity. 

Due to the azimuthally anisotropic spatial density distribution in semi-central heavy-ion collisions, hard partons produced in rare collisions, tend to experience azimuthally asymmetric energy loss~\cite{Wang:2003mm,Majumder:2006we,Bass:2008rv}. 
This leads to an azimuthal anisotropy in both high transverse momentum (high $p_T$) hadron~\cite{ATLAS:2018ezv} and jet distributions~\cite{ATLAS:2013ssy,ALICE:2015efi}.
This high $p_T$ anisotropy, goes hand-in-hand with the angle-integrated, centrality-dependent suppression of the high $p_T$ jet and hadron spectrum in heavy-ion collisions (with greater  suppression appearing in the more central collisions)~\cite{ATLAS:2012tjt,CMS:2011iwn,ALICE:2013dpt,ATLAS:2015qmb,CMS:2012aa,ALICE:2010yje,STAR:2003fka,PHENIX:2001hpc, Burke:2013yra, Majumder:2010qh,Cao:2024pxc}. 

Simulations of jet energy loss in a fluid medium~\cite{Noronha-Hostler:2016eow,Kumar:2019uvu,He:2022evt} compute and obtain both the angle integrated and angle dependent suppression by changing the underlying space-time fluid profile used to compute the local jet transport coefficients~\cite{Baier:2002tc, Majumder:2012sh, Kumar:2020wvb, Majumder:2008zg}. The underlying fluid profile is obtained by rigorous comparison of bulk evolution simulations with experimental data~\cite{Song:2010mg,Shen:2014vra,Shen:2014vra, Bernhard:2019bmu, Moreland:2018gsh}.

Recently, measurements of $p$-$p$ and $p$-$Pb$ collision events (referred to as small systems), with high multiplicity ($N\gtrsim 90$), yield a variety of azimuthal anisotropies~\cite{ATLAS:2017hap,ALICE:2014dwt,CMS:2015yux} in the soft bulk sector ($p_T \!\lesssim\! 3$~GeV). This indicates the possible formation of a QGP in these small systems.
There have been several theoretical efforts to understand the origin of the low-$p_T$ azimuthal anisotropies in small systems~\cite{Dumitru:2010iy,Kovner:2012jm,Gyulassy:2014cfa,Lappi:2015vta,Altinoluk:2015uaa,Dumitru:2015gaa,Altinoluk:2016vax,Hagiwara:2017ofm,Altinoluk:2018ogz,Altinoluk:2020psk,Hatta:2020bgy}.
The corroborating evidence from high $p_T$ probes is, however, inconclusive.
Experimental data on hadrons at large $p_T$ show a significant azimuthal anisotropy~\cite{ATLAS:2016yzd,ATLAS:2019vcm}, in both $p$-$p$ (measured up to 10~GeV), and $p$-$Pb$ (measured up to 50~GeV) collisions.
However, there is no evidence of suppression in the angle integrated spectrum in $p$-$Pb$ collisions (at high-$p_T \gtrsim 5$~GeV)\cite{CMS:2016xef,ALICE:2015umm,ATLAS:2022kqu}.

The origin of an azimuthal anisotropy at high-$p_T$, in the absence of any suppression of the angle integrated spectrum in $p$-$A$, can be found in the presence of the same anisotropy in $p$-$p$ collisions themselves. In the current effort, this anisotropy will arise due to the existence of a small transverse momentum in the initial state: Partons in the proton carry a small transverse momentum $k_\perp$, in addition to the large longitudinal momentum, which is a fraction ($x$) of the proton's momentum, i.e. $p^+ = x P^+$. 

This small transverse momentum will lead to azimuthal anisotropies in both the polarization dependent and polarization averaged cross sections. In this Paper, we study how  transverse momentum dependent parton distribution functions (TMDPDFs) and fragmentation functions (TMDFFs)~\cite{Collins:1981uw}, can generate azimuthal anisotropy at high $p_T$ in $p$-$p$ and $p$-$Pb$ collisions.
In particular, we find that the combination of the Boer-Mulders' process of polarized partons from unpolarized hadrons~\cite{Boer:1997nt} and the Collins' process~\cite{Collins:1992kk} of polarized parton fragmentation into unpolarized hadrons lead to a large azimuthal anisotropy.
The small transverse momentum $k_\perp$ in $p$-$p$ will be enhanced by a factor of $A^{1/3}$ (with $A$ the mass number of the nucleus) in $p$-$A$ collisions, to account for the possibility of soft pre-scattering of the hard parton from the proton within the nucleus, prior to the hard scattering that leads to jet formation.

Due to the small system size, momentum conservation in the initial state becomes more apparent in the particle distributions from the final state in $p$-$p$ and $p$-$A$: The transverse momentum $k_\perp$ of the hard parton, that will participate in hard scattering is balanced by the remainder of the proton. 
As a result, the net transverse momentum of the produced dijet system is partially balanced by the net transverse momentum of the bulk system.
Simulations using the PYTHIA generator~\cite{Bierlich:2022pfr} will be used to show that an intrinsic transverse momentum will yield both greater particle production over a range of rapidities, as well as the correlation between the dijet pair momentum ($\bm{q}_T = \bm{p}_{jet\, 1} + \bm{p}_{jet\, 2}$) and the event plane angle(s). Momentum conservation will correlate the $\cos(n\phi)$ asymmetries of the hard sector with the event plane of the bulk system, leading to corresponding $v_n$'s in the hard sector. PYTHIA simulations, currently do not include polarization dependent matrix elements. Thus, our final results will be semi-analytic, assuming that the dijet pair angle is completely correlated with the event plane angle(s).

The paper is organized as follows: In Sec.~\ref{sec:Theory}, we present the theoretical framework of polarization dependent and polarization averaged scattering with TMDPDFs and TMDFFs. We also present the matrix elements for the processes that will be used. In Sec.~\ref{Pheno-v2}, we present a phenomenological description of TMDPDF and TMDFFs that will be used to numerically evaluate these quantities. 
In Sec.~\ref{sec:simulations}, we present the overarching phenomenological picture of how an elliptic anisotropy can originate in a small system, in the presence of hard scattering.
We also carry out a modified PYTHIA simulation to demonstrate this picture and correlate the event plane angle(s) with the direction of the net momentum of the dijets ($\bm{q}_T = \bm{p}_{jet\, 1} + \bm{p}_{jet\, 2}$).
In Sec.~\ref{sec:Results}, we present our results for the cross section and the azimuthal anisotropy, and compare with data from the ALICE and ATLAS collaborations. 
A summary and outlook is presented in Sec.~\ref{sec:Conclusion}.

\section{Theoretical Framework}\label{sec:Theory}

In the standard collinear factorization of hard processes~\cite{Collins:1985ue}, the outgoing hadron distributions are azimuthally symmetric. The two incoming partons that participate in the hard scattering, approach head-to-head, perfectly collinearly, and leave the hard scattering back-to-back.  There is no preferred direction in the azimuthal plane of the initial state.
Instead, if one were to include the more realistic transverse momentum dependent (TMD-) PDFs~\cite{Boer:2010zf,Pisano:2013cya,denDunnen:2014kjo,Boer:2014tka}, and FFs~\cite{Bacchetta:1999kz,Anselmino:2004ky,Anselmino:2005sh,DAlesio:2004eso}, the two incoming partons ($a,b$) are no longer perfectly collinear. Each has a transverse momentum ($\bm{k}^a_\perp, \bm{k}^b_\perp$). As a result, the direction $\bm{k}^a_\perp + \bm{k}^b_\perp$ breaks the azimuthal symmetry of the process, resulting in momentum correlations in the transverse plane of the hadronic scattering.

Following~\cite{Anselmino:2004ky,Anselmino:2005sh}, we assume that at high energies the momentum hierarchy between the longitudinal momenta and the transverse momenta allows the pion production cross section to factorize into a soft parton correlator and a hard scattering matrix element 
\begin{align}
&\frac{d\sigma}{dy d^2 P_{T}}
=\sum_{a,b,c,d}\int \frac{dx_a dx_b dz}{2 \pi^2 z^3 s} d^2k_{\bot a} d^2 k_{\bot b} d^3 k_{\bot C}  \nonumber\\
&\delta(\bm{k}_{\bot C}\cdot \hat{p}_c)J(\bm{k}_{\bot C}) \Gamma^{\sigma\mu}(x_a,k_{\bot a}) \Gamma^{\alpha \nu}(x_b, k_{\bot b})  \nonumber\\
&\hat{M}_{\mu\nu\rho} \left(\hat{M}_{\sigma\alpha\beta}\right)^*  \Delta^{\rho\beta}(z,k_{\bot C}) \delta(\hat{s} + \hat{t} + \hat{u})\;, \label{eq:cross-section}
\end{align}
where $J(\bm{k}_{\bot C}) = \frac{(E_C^2+\sqrt{\bm{p}_C^2-\bm{k}_{\bot C}^2})^2}{4(\bm{p}_C^2-\bm{k}_{\bot C}^2)}$ and the Mandelstamn variables of the partonic system are given by $(\hat{s},\hat{t},\hat{u})$.
In the equation above, $\hat{M}_{\mu\nu\rho} \left(\hat{M}_{\sigma\alpha\beta}\right)^*$ is the hard partonic squared amplitude, $\Gamma$ are the initial state correlators and $\Delta$ are the final state fragmentation correlator.

The leading twist gluon generalized PDF reads \cite{Mulders:2000sh}
\begin{align}
&\Gamma^{\mu\nu}_{P}(x,k_{\bot})
= \nonumber\\
&\int \frac{d(\xi\cdot P) d^2\xi_T}{(2\pi)^3}~ e^{i(p\cdot \xi)}
\frac{\langle P | {\rm Tr}\left[F^{\mu\rho}(0)F^{\nu\sigma}(\xi)\right] | P \rangle n_{\rho}n_{\sigma}}{(p\cdot n)^2}
\nonumber\\
&= \frac{
	-g_T^{\mu\nu} f_g(x,k_{\bot}) 
+ \left(\frac{k_{\bot}^{\mu}k_{\bot}^{\nu}}{M_p^2} + g_T^{\mu\nu}\frac{k_{\bot}^2}{2M_p^2} \right)h_g^{\bot}(x,k_{\bot}^2)}{2x}\;,
\end{align}
where $f(x, k_{\bot})$ is the unpolarized gluon distribution.
The gluon Boer-Mulders' function $h^{\bot}(x,k_{\bot}^2)$ can be interpreted as the distribution of linearly polarized gluons in the nucleon.
Using the following gluon polarization vectors,
\begin{align}
\epsilon^\mu_{\lambda_1} 
=& \left(0,-\lambda_1 e^{-i\phi},-i\lambda_1 e^{i\phi},0\right)\;,\;,\\
\epsilon^{\mu*}_{\lambda_2} 
=& \left(0,-\lambda_2 e^{-i\phi},-i\lambda_2 e^{i\phi},0\right)\;,
\end{align}
we can project the gluon generalized PDF onto gluon helicity states
\begin{align}
&\Gamma^{\lambda_1, \lambda_2}_{P}(x,k_{\bot})
= \nonumber\\
& \frac{
	-\delta^{\lambda_1,\lambda_2} f_g(x,k_{\bot}) 
+ \frac{k_T^2}{2M_p^2} \delta^{\lambda_1,-\lambda_2} h_g^{\bot}(x,k_{\bot}^2)}{2x}\;.\label{eq:CorrelatorPDF}
\end{align}

The quark generalized PDF is given by
\begin{align}
&\Gamma^{\mu\nu}_{P}(x,k_{\bot})
= \nonumber\\
&\int \frac{d(\xi\cdot P) d^2\xi_T}{(2\pi)^3}~ e^{i(p\cdot \xi)}
\frac{\langle P | \bar{\psi}(0) \psi(\xi) | P \rangle n^{\mu}n^{\nu}}{(p\cdot n)^2}
\nonumber\\
&= \frac{1}{2}\left[\slashed{P} f_q(x,k_{\bot}) + h_q^{\bot}(x, k_{\bot}) \frac{\sigma_{\mu\nu} k_{\bot}^{\nu} P^{\nu}}{2M_p}\right]\;,
\label{eq:qCorrelator}
\end{align}
where $\sigma_{\mu\nu}=[\gamma_\mu,\gamma_\nu]$.
Similar to the gluon, the distribution of unpolarized quarks is given by $f_q(x,k_{\bot})$.
The time-reversal odd function $h_q^{\bot}(x, k_{\bot})$ is the quark Boer-Mulders' function, which can be interpreted as the distribution of transversely polarized quarks in an unpolarized proton.

Analogously, the fragmentation correlators are written
\begin{align}\label{eq:CorrelatorFF}
&\Delta_g^{\lambda_1,\lambda_2} (z,k_{\bot}) 
= \nonumber\\
&\frac{
-\delta^{\lambda_1,\lambda_2} D_g(z,k_{\bot}^2) 
+\delta^{\lambda_1,-\lambda_2}\frac{k_{\bot}^2}{2M_{\pi}^2} H_g^{\bot}(z,k_{\bot}^2)}{2/z},\\
&\Delta_q (z,k_{\bot}) 
	= \frac{
		\slashed{p} D_q(z,k_{\bot}^2)
+\frac{\sigma_{\mu\nu} k_{\bot}^{\nu} P^{\nu}}{2M_p} H^{\bot}_q(z,k_{\bot}^2)}{2/z}.
\end{align}
The TMD fragmentation of unpolarized gluons or quarks to unpolarized pions with momentum $z\bm{p}_c + \bm{k}_{\bot C}$ is given by $D_{g/q}(z,k_{\bot}^2)$.
The equivalent to the BM function is the Collins' function $H_{g/q}^{\bot}(z,k_{\bot}^2)$, which describes the fragmentation of a transversely polarized gluon or transversely polarized quarks to an unpolarized pion \cite{Collins:1992kk}.

\subsection{Unpolarized partons cross section}
For the case of unpolarized partons, the cross section from Eq.~(\ref{eq:cross-section}) can be simplified further by setting the same helicities between the matrix element and its complex conjugate, leading to
\begin{align}
&\frac{d\sigma}{dyd^2P_T}
= \sum_{a,b,c,d}\int \frac{dx_a dx_b dz}{16 \pi^2 x_a x_b z^2 s} d^2k_{\bot a} d^2 k_{\bot b} d^3 k_{\bot C} \nonumber\\
&\delta(\bm{k}_{\bot C}\cdot \hat{p}_c)J(\bm{k}_{\bot C})
f^{a/A}\left(x_a, {k}_{\bot a}\right) f^{b/B}\left(x_b, {k}_{\bot b} \right)\nonumber\\
&|\hat{M}_{ab}^{cd}|^2
D^{C/c}\left(z,{k}_{\bot C} \right)
\delta(\hat{s} + \hat{t} + \hat{u})\;, \label{eq:Unpolarized-cross-section}
\end{align}
where $|\hat{M}_{ab}^{cd}|^2$ is now the usual unpolarized partonic squared matrix element that can be found in the literature \cite{Field:1989uq}.

Due to the transverse momenta of the partons, the partonic center of mass frame $(\bm{q} = \bm{k}_a + \bm{k}_b)$ is off-center in the transverse plane of the hadronic scattering.
This leads to transverse momentum correlations between the final state partons.
Taking the limit of small transverse momentum $k_{\bot a}\ll p_a$, the incoming partonic momenta, with a small component in the $x$-direction, can be written $\p_a=p_a(1,\sin\theta_a,0,\cos\theta_a)$, where $\theta_a$ is the polar angle in the hadronic center of mass frame.
Consider a typical partonic matrix element squared, where the outgoing parton $c$ is produced at mid-rapidity ($\theta_c=\pi/2$), along an azimuthal angle $\phi_c$, 
\begin{align}
\mathcal{M}
\propto \frac{\hat{s}}{\hat{t}}
\simeq \frac{-\hat{s}}{2p_a p_c (1-\theta_a \cos(\phi_c))}\;.
\end{align}
By expanding the denominator in the small angle limit $\theta_a \ll 1$, we obtain
\begin{align}
\mathcal{M}
&\simeq \frac{-\hat{s}}{4p_a p_c }\left[2 + \theta_a^2 + 2\theta_a \cos(\phi_c) + \theta_a^2 \cos(2\phi_c)\right]\;.
\end{align}
These azimuthal correlations generated by the intrinsic transverse momentum are suppressed by powers of the large transverse momentum, as $\theta_a = \sfrac{k_{\bot a}}{p_a}$. Thus, while azimuthal anisotropies do arise in unpolarized cross sections, they are suppressed by powers of the hard scale, and have a diminishing contribution at high-$p_T$.

\subsection{Polarized partonic cross section}
Since polarized distributions change the helicity between the matrix elements and its complex conjugate (c.f. Eqs.~(\ref{eq:CorrelatorPDF}-\ref{eq:CorrelatorFF})), inclusive unpolarized pion production requires two polarized distributions of the same parton species.
Either both initial partons are polarized or one initial parton and one final parton are polarized, which we will refer to as $\rm BM\otimes BM$ or $\rm BM \otimes C$ effects, respectively.
In the following four sections we will present the relevant contributions for gluon initiated processes and quark initiated processes.

\subsubsection{Gluon initiated: Boer-Mulders' Collins' effect}
We start with the $\rm BM \otimes C$ effect, where a polarized gluon, stemming from a Boer-Mulders' function, scatters with an unpolarized gluon, producing a polarized gluon that fragments to an unpolarized pion, with a Collins' function.
Here, the matrix element times the PDFs and FF is written as,
\begin{align}
&\Sigma^{\rm BM\otimes C}
= {\rm Re} \sum_{\lambda=\pm1}  H^{\bot (1)}_{C/g}(z,k_{\bot C})
\nonumber\\
&\times \left[h^{\bot (1)}_{g/A}(x_a,k_{\bot a}^2)f_{g/B}(x_b,k_{\bot b}^2)\hat{M}_{\lambda,\lambda}^{\lambda,\lambda} \left(\hat{M}_{-\lambda,\lambda}^{-\lambda,\lambda}\right)^*\right.\nonumber\\
&\left.+ f_{g/A}(x_a,k_{\bot a}^2)h^{\bot (1)}_{g/B}(x_b,k_{\bot b}^2)\hat{M}_{\lambda,\lambda}^{\lambda,\lambda} \left(\hat{M}_{\lambda,-\lambda}^{-\lambda,\lambda}\right)^*\right]  \;. \label{eq:sigmaBMC}
\end{align}
We list the Matrix element times the PDFs and FF for gluon-gluon
\begin{align}
\mathcal{T}^{\rm BM \otimes C}_{gg\to gg}
=& g_s^4 \frac{N^2}{N^2-1} \frac{t^2 + tu + u^2}{t^2} e^{-4\lambda i(\phi_{12}-\phi_{23})}\;,
\end{align}
where using the momentum angle $\p=(p,\theta,\phi)$, we write
\begin{align}
\tan\phi_{ij} = \tan\frac{\phi_j-\phi_i}{2} \frac{\sin\frac{\theta_j-\theta_i}{2}}{\sin\frac{\theta_j+\theta_i}{2}}\;.
\end{align}
After performing the helicity sums, the contribution to the cross section becomes
\begin{align}
&\Sigma^{\rm BM\otimes C}_{gg\to gg}
=2\hat{D}^{C/g}(z,k_{\bot C})
\nonumber\\
&
\Bigl[ h^{\bot (1)}_{g/A}(x_a,k_{\bot a}^2)f_{g/B}(x_b,k_{\bot b}^2)
	\hat{M}^0_1 \hat{M}^0_2 \cos\left(4(\phi_{12} - \phi_{23}) \right)
\nonumber\\
&
+ f_{g/A}(x_a,k_{\bot a}^2)h^{\bot (1)}_{g/B}(x_b,k_{\bot b}^2)
\hat{M}^0_1 \hat{M}^0_3 \cos\left(4(\phi_{12} - \phi_{13}) \right) \Bigr].
\end{align}
We define matrix elements stripped of their angular correlations,
\begin{align}
&\hat{M}^0_1\hat{M}^0_2
= \frac{9}{2} g_s^4 \left[1+\frac{u}{t} + \frac{u^2}{t^2}\right]\;,\\
&\hat{M}^0_1\hat{M}^0_3
= \frac{9}{2} g_s^4 \left[1+\frac{t}{u} + \frac{t^2}{u^2}\right]\;.
\end{align}

\subsubsection{Gluon initiated: Double Boer-Mulders' effect}
Another contribution to the cross section comes from the scattering of two polarized gluons leading to unpolarized gluons in the final state.
The matrix element times the PDFs and FF is now written
\begin{align}
\Sigma^{\rm BM\otimes BM}_{gg\to gg}
=& \sum_{\lambda=\pm1} 
h^{\bot (1)}_{g/A}(x_a,k_{\bot a}^2)
h^{\bot (1)}_{g/B}(x_b,k_{\bot b}^2)
\nonumber\\
	&
	\hat{M}_{\lambda,\lambda}^{\lambda,\lambda} 
	\left(\hat{M}_{-\lambda,-\lambda}^{\lambda,\lambda}\right)^*
	D_{C/g}(z,k_{\bot C})
	\;,
\end{align}
with the following matrix element 
\begin{align}
\mathcal{T}^{\rm BM \otimes BM}_{gg\to gg}
=& g_s^4 \frac{N^2}{N^2-1} \frac{t^2 + tu + u^2}{s^2} e^{-4\lambda i(\phi_{23}-\phi_{13})}\;.
\end{align}

Performing the helicity sums, yields
\begin{align}
&\Sigma^{\rm BM\otimes BM}_{gg\to gg}
=4
h^{\bot (1)}_{g/A}(x_a,k_{\bot a}^2)
h^{\bot (1)}_{g/B}(x_b,k_{\bot b}^2)
D_{C/g}(z,k_{\bot C})
\nonumber\\
&
\hat{M}^0_2 \hat{M}^0_3
	\cos\left(4(\phi_{13} - \phi_{23})\right)
\;,
\end{align}
with 
\begin{align}
\hat{M}^0_2\hat{M}^0_3
= \frac{9}{2} g_s^4 \left[\frac{t^2 + tu + u^2}{s^2}\right]\;.
\end{align}
\subsubsection{Quark initiated: Boer-Mulders' Collins' effect}
For the quark-quark scattering, the square matrix element of the $BM \otimes C$ effect lead to odd number of Dirac matrices and only the interference of the $t$- and $u$-channel diagrams contribute to the cross section.
The matrix element times the PDFs and FF is written
\begin{align}
&\Sigma^{\rm BM\otimes C}_{qq \to qq}
=2H^{\bot (1)}_{C/q}(z,k_{\bot C})
\biggl[ \hat{M}^0_1 \hat{M}^0_2 h^{\bot (1)}_{q/A}(x_a, k_{\bot a})
\nonumber\\
&
\times f_{q/B}(x_b, k_{\bot b})
+ \hat{M}^0_1 \hat{M}^0_3 f_{q/A}(x_a, k_{\bot a}) h^{\bot (1)}_{q/B}(x_b, k_{\bot b}) \biggr].
\end{align}
The matrix elements are given by the trace of ten Dirac matrices which evaluates to
\begin{align}
\hat{M}^0_1\hat{M}^0_2
=&-\frac{8}{27} \frac{1}{tu}
\biggl[
t\left(\bm{k}_{\bot a} \cdot \bm{p}_2\right)\left(\bm{k}_{\bot C} \cdot \bm{p}_4\right)
\nonumber\\
&+t\left(\bm{k}_{\bot a} \cdot \bm{p}_2\right)\left(\bm{k}_{\bot C} \cdot \bm{p}_2\right)
-s u (\bm{k}_{\bot a} \cdot \bm{k}_{\bot C})
\biggr]\;,\\
\hat{M}^0_1\hat{M}^0_3
=&-\frac{8}{27} \frac{1}{tu}
\biggl[
u\left(\bm{k}_{\bot b} \cdot \bm{p}_1\right)\left(\bm{k}_{\bot C} \cdot \bm{p}_4\right)
\nonumber\\
&+u\left(\bm{k}_{\bot b} \cdot \bm{p}_1\right)\left(\bm{k}_{\bot C} \cdot \bm{p}_1\right)
-s t (\bm{k}_{\bot b} \cdot \bm{k}_{\bot C})
\biggr]\;.
\end{align}
The scalar products above lead to an azimuthal correlation of the form $\cos(\phi_{\bm{k}_{\bot a}})$ and $\cos(\phi_{\bm{k}_{\bot b}})$, which produces a contribution to the $v_1$ in the azimuthal distribution (c.f. Appendix~\ref{app:QuarkScattering}).

\subsubsection{Quark initiated: Double Boer-Mulders' effect}
Similarly, the quark $BM \otimes BM$ effect is given by the interference of the $t$- and $u$-channel diagrams which leads to the following Dirac trace
\begin{align}
\Sigma^{\rm BM\otimes C}_{qq \to qq}
=&2 D^{C/q}(z,k_{\bot C})h^{\bot (1)}_{q/A}(x_a, k_{\bot a})h^{\bot (1)}_{q/B}(x_b, k_{\bot b})
\nonumber\\
& \times\hat{M}^0_2 \hat{M}^0_3 \;,
\end{align}
with the matrix element
\begin{align}
\hat{M}^0_2 \hat{M}^0_3
=&-\frac{8}{27} \frac{1}{tu}
\biggl[
s\left(\bm{k}_{\bot a} \cdot \bm{p}_3\right)\left(\bm{k}_{\bot b} \cdot \bm{p}_4\right)
\nonumber\\
	&+s\left(\bm{k}_{\bot a} \cdot \bm{p}_4\right)\left(\bm{k}_{\bot b} \cdot \bm{p}_3\right)
+t u (\bm{k}_{\bot a} \cdot \bm{k}_{\bot b})
\biggr]\;.
\end{align}
Here the scalar products lead to azimuthal correlation of the form $\cos(\phi_{\bm{k}_{\bot a}})\times\cos(\phi_{\bm{k}_{\bot b}})$ which produces a contribution to the $v_2$ in the azimuthal distribution (c.f. \ref{app:QuarkScattering}).

All the matrix elements listed above for quark and gluon scattering will have to be convoluted with the appropriate TMDPDF and TMDFF, according to Eqs.~\eqref{eq:cross-section}. These non-perturbative functions, typically obtained from fits to data from other experiments, are not well constrained in many cases. In the subsequent section, we will describe how these functions are parameterized and how we modify these for use in $p$-$A$ collisions. Final numerical results are presented in Sec.~\ref{sec:Results}.

\section{Phenomenology of TMDPDF and TMDFF}
\label{Pheno-v2}

In the preceding section, we outlined the matrix elements of all the processes that could contribute to high-$p_T$ hadron production in $p$-$p$ (and $p$-$A$) collisions. These include both polarization averaged and polarization dependent contributions. 
To complete the process, these matrix elements have to be convoluted with the appropriate TMDPDF and TMDFF. 

In this section, we will discuss our chosen parameterizations for both the polarization averaged and polarization dependent PDF and FF. It will come as no surprise that some of these functions are not well constrained by experimental data. Our choice of parameterization is further constrained by the fact that our parameterization also has to be easily modified to the case of $p$-$A$. As implemented below, this will force us to use factorized ($x$ and $k_\perp$) distributions. 

\subsection{Mean transverse momentum}
\begin{figure}
\begin{center}
	\includegraphics[width=0.45\textwidth]{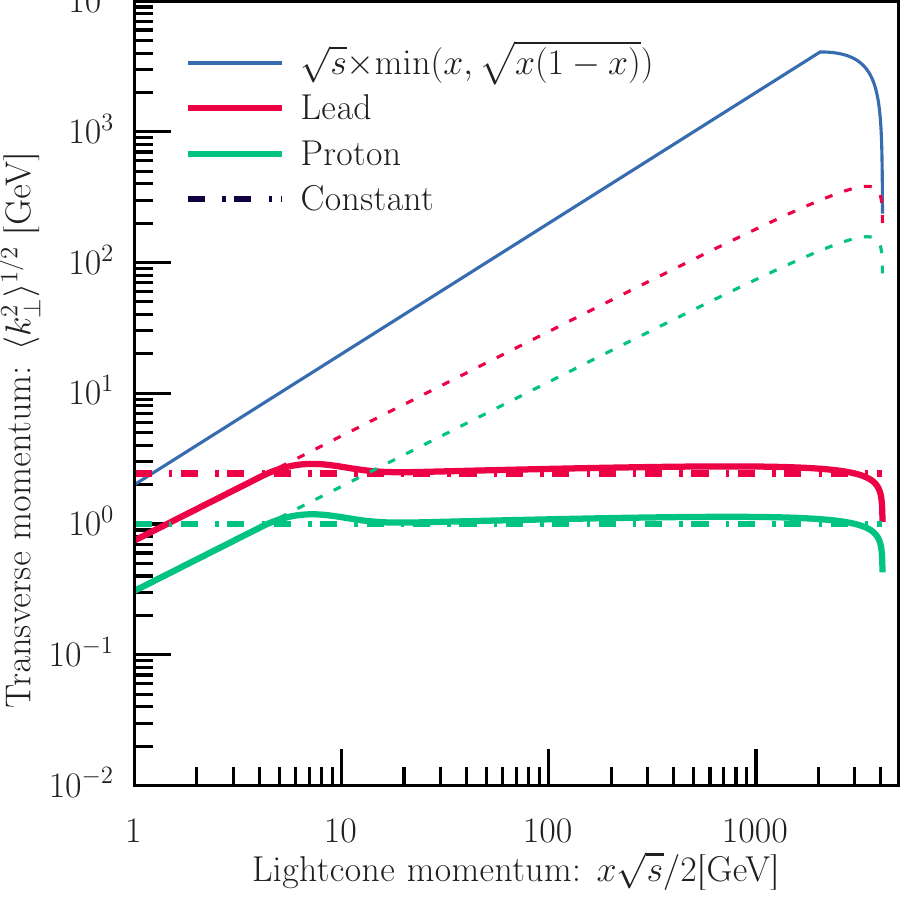}
\end{center}
\caption{Different models of the mean transverse momentum $\langle k_\perp^2 \rangle$ for $p$-$p$ and $p$-$A$ collisions where one side is enhanced by $A^{1/3}$. The dot-dashed lines indicate the case of $x$-independent $\langle k_\perp^2 \rangle = A^{1/3}$ GeV. Solid lines indicate the $x$-dependent model of Eq.~\eqref{eq:kTFct}. The dotted lines indicate what would happen to this model if the model of $x<x_0$ is allowed to apply for the entire range of $x$. The solid blue line is the hard kinematic bound in Eq.~(\ref{eq:KinematicConstraint}). }
\label{fig:kTFct} 
\end{figure}

\begin{figure}
\includegraphics[width=0.45\textwidth,angle=90]{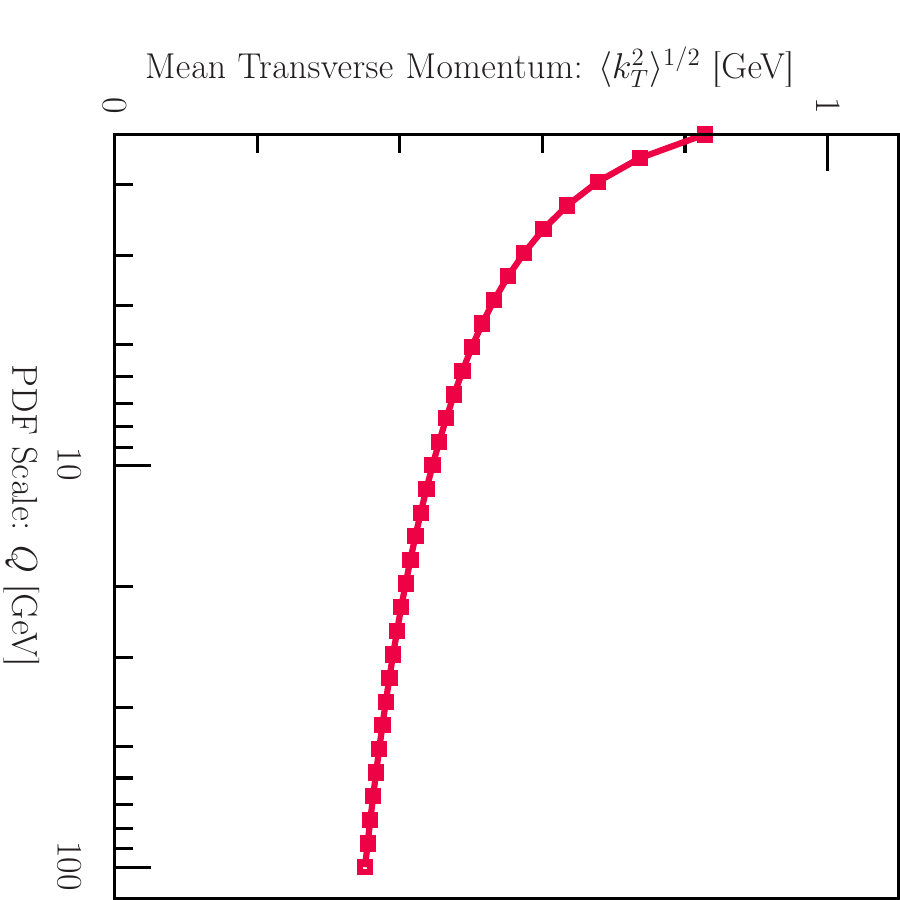}
\caption{Mean transverse momentum as a function of the hard scale $Q$, obtained by carrying out the averaging of $\langle k_\perp^2 \rangle (x,Q^2)$, over the allowed range of $x$ as described in Eq.~\eqref{eq:MeankT}. Due to the shift of parton distribution functions to smaller $x$ at larger $Q$, the mean transverse momentum gradually reduces with increasing $Q$. }
\label{fig:MeankT}
\end{figure}

There have been several experimental and Lattice QCD results that have enabled multi-dimensional parameterizations of TMDPDFs [$f(x,k_\perp^2,Q^2)$, where the $x$ and $k_\perp$ dependence is not factorized]~\cite{BermudezMartinez:2018fsv,BermudezMartinez:2021lxz}.
However, there are no parameterizations of un-factorized TMDPDFs including nuclear effects, that can be used for $p$-$A$ collisions. There are two specific effects that concern us, the modification of the $x$ distribution of partons in the nucleus $A$, referred to as shadowing~\cite{Arnold:1983mw}, and the modification of the transverse momentum of a parton in the $p$ undergoing pre-scattering in the nucleus prior to the hard exchange that leads to dijets, referred to as the Cronin effect~\cite{Cronin:1974zm}. In the absence of a  pre-scattering formalism that can simultaneously describe both these effects, we are left with a factorized parameterization of both these effects.

In the literature, for the case of factorized distributions, the transverse momentum dependence is typically approximated by a Gaussian \emph{Ansatz}~\cite{Gamberg:2005ip,DAlesio:2004eso}
\begin{align}
f(x,k_{\bot}, Q^2)
=& \frac{4\pi}{\langle k_\bot^2\rangle} e^{-\frac{k_\bot^2}{\langle k_\bot^2\rangle}} f(x,Q^2)\;.
\end{align}
In the presence of a large nucleus, partons may undergo scattering off multiple nucleons before the hard exchange that leads to the energetic dijets.
For the case of polarization averaged processes, it has been demonstrated that the extra acquired mean squared transverse momentum in $p$-$A$ collisions($Q_{\perp,A}^2$) is  proportional to the length traversed, i.e.,  $Q_{\perp,A}^2 \propto A^{1/3} $~\cite{Guo:1997it, Fries:2002mu, Majumder:2007hx, Kovchegov:1999yj, Mueller:1989st, Gribov:1983ivg}. Thus the total mean transverse momentum from pre-scattering, prior to the hard interaction may be expressed as
\begin{eqnarray}
\langle k_\perp^2 \rangle_{pA} \simeq Q_{\perp,A}^2 + \langle k_\perp^2 \rangle_{pp}  \; .
\end{eqnarray}
In the equation above, $\langle k_\perp^2 \rangle_{pp}$ is the mean transverse momentum from the TMDPDF of the proton prior to interaction with the nucleus. In this first attempt to calculate the multiple moments of the elliptic anisotropy, we will invoke the \emph{Ansatz}:
\begin{align}
    Q_{\perp,A}^2 \sim A^{1/3} \langle k_\perp^2 \rangle_{pp} \gg \langle k_\perp^2 \rangle_{pp}.
\end{align}
There are two aspects to this: The stronger statement is that the transverse momentum accrued from scattering off each nucleon is nearly equal to the intrinsic transverse momentum of typical partons in a nucleon, and that the nucleus in the $p$-$A$ case is large, i.e., $A^{1/3} \gg 1$.

Given the above assumptions, for calculational convenience, we can incorporate the extra transverse momentum from pre-scattering within an effective transverse momentum distribution of the parton (from the proton) before hard scattering in $p$-$A$. Thus for the case of $p$-$A$, the parton distribution function in the proton is modified as:
\begin{align}
    f_{pA}(x,k_{\bot}, Q^2)
	=& \frac{4\pi}{A^{1/3}\langle k_\bot^2\rangle_{pp}} e^{-\frac{k_\bot^2}{A^{1/3}\langle k_\bot^2\rangle_{pp}}} f_{pA}(x,Q^2)\;.
\end{align}
In the above equation, we will employ an nPDF $f_{pA}(x,Q^2)$ for the $A$ ion side.

In order to make sure the partons are generally collinear with the proton, we impose a kinematic constraint on the transverse momentum of the partons \cite{Anselmino:2005sh}
\begin{align}\label{eq:KinematicConstraint}
k_{\bot} \leq \sqrt{s} \times {\rm min}\left(x,\sqrt{x(1-x)}\right)\;.
\end{align}
While the factorization of the transverse momentum simplifies the calculation significantly, due to kinematic constraints the low-$x$ partons will not contribute to the cross section.

In an attempt to capture the complex relation between the transverse momentum and the momentum fraction, we use an $x$-dependent $\langle k_\bot^2 \rangle$ \emph{ansatz}.
A parameterization of the mean transverse momentum in $p$-$p$ collisions was presented in Ref.~\cite{DAlesio:2004eso} as follows
\begin{align}
\langle k_\bot^2 \rangle^{1/2} (x) 
= \langle k_\bot^2 \rangle_0^{1/2} x^{0.8} (1-x)^{0.15}\;.
\end{align}
However, this \emph{ansatz} can become large at intermediate $x$ as shown by the doted lines in Fig.~\ref{fig:kTFct}.

Multiple scattering in deconfined matter leads to transverse momentum broadening governed by the transport coefficient $\hat{q}$~\cite{Arnold:2008vd}.
The mean transverse momentum is proportional to $\sqrt{\hat{q} L}$, where $L$ is the length of the medium.
Perturbatively, $\hat{q}$ is governed by a logarithmic dependence $(\log E/E_0)$ on the energy of the hard parton $E$ and a soft screening scale $E_0$~\cite{Arnold:2008vd}.
Guided by this result, we modify the $x$-dependent \emph{ansatz} of Ref.~\cite{DAlesio:2004eso} as follows\footnote{For momentum fractions in the range $x_0 \leq x \leq 10 x_0$, we use a hyperbolic tangent $\left(\tanh\frac{x-x_0}{x_0}\right)$ to smoothly match the logarithmic behavior to the power-law.},
\begin{eqnarray}\label{eq:kTFct}
&\langle k_\bot^2 \rangle^{1/2} (x) 
= \langle k_\bot^2 \rangle_0^{1/2} \left((1-x)\sqrt{\frac{s}{s_0}}\right)^{0.15} A^{1/6} \nonumber\\
&\times
\begin{cases}
\left(\frac{x}{x_0}\sqrt{\frac{s}{s_0}}\right)^{0.8}\;, &\text{ if } x \leq x_0\\
\left[1 + b_0 \log\left(\frac{x}{x_0}\sqrt{\frac{s}{s_0}}\right)\right]^{1/2} \;, &\text{ if } x > x_0
\end{cases}
\end{eqnarray}
where we take $\langle k_\bot^2 \rangle_0 = 0.9$~GeV$^2$, $x_0 = 10^{-3}$, $\sqrt{s_0}=8.16$~TeV and $b_0=0.1$.
This is indicated by the solid red line in Fig.~\ref{fig:kTFct}.
Using these constants one can reproduce the distribution of Ref.~\cite{DAlesio:2004eso} at $x \leq x_0$.
Throughout this manuscript, unless stated otherwise, we will use this $x$-dependent ansatz, we refer to~\cite{Soudi:2023epi} for results using a constant $\langle k_\bot^2 \rangle$.

Since Eq.~(\ref{eq:kTFct}) involves non-trivial $x$ dependence, the true mean transverse momentum of the PDF is not directly the value of $\langle k_\bot^2 \rangle^{1/2} (x)$.
To obtain what will refer to as the effective mean transverse momentum, one needs to integrate the full PDF as follows
\begin{align}\label{eq:MeankT}
\langle k_T^2\rangle(Q) = \frac{\int dx_a \langle k_\bot^2\rangle (x_a) f(x_a,Q)}{\int dx_a f(x_a,Q)}\;,
\end{align}
where $Q$ is the hard scale of the PDF.
In contrast to Fig.~\ref{fig:kTFct}, the curve we obtain in Fig.~\ref{fig:MeankT} leads to a mean transverse momentum lower than $1$~GeV for all scales.
This is because the support of the PDF becomes more and more concentrated at low $x$ as the scale increases. We point out these TMDPDF's, even those for the case of $p$-$p$ collisions, are phenomenological, with coefficients dialed to fit data. These do not obey the TMD evolution equations~\cite{Collins_2011}.

\subsection{Fragmentation Functions}
The theoretical description of the production of pions from the hadronic scatterings, requires fragmentation functions as described in Eq.~(\ref{eq:CorrelatorFF}).
Collinear and TMD unpolarized fragmentation functions have been studied extensively and are well constrained by data \cite{Kniehl:2000fe,Aybat:2011zv,Barry:2023qqh}.
In order to simplify the analysis, we will employ the same factorized Gaussian \emph{Ansatz} for the TMD fragmentation functions, as we have done for PDFs, i.e. 
\begin{align}
	D_{C/c}(z,k_{\bot C}^2) = \frac{4\pi}{\langle k_{\bot C}^2 \rangle} e^{-\frac{k_{\bot C}^2}{\langle k_{\bot C}^2 \rangle}} D_{C/c}(z)\;.
\end{align}

One can also consider a $z$-dependent $\langle k_{\bot C}^2 \rangle$ \emph{ansatz} for the fragmentation functions \cite{DAlesio:2004eso}. However, this plays a very minor role for the processes that are being studied.
Since the pion transverse momentum $\bm{k}_{\bot C}$ is transverse to the momentum of the parton $c$, the fragmentation functions will not contribute to the azimuthal anisotropies in the plane transverse to the beam axis.
For the $z$-dependence of the fragmentation functions, we will use the KKP parameterization \cite{Kniehl:2000fe}. Similar to the case of the TMDPDF's, these TMDFF's are phenomenological and do not obey TMD evolution~\cite{Collins_2011}.

\subsection{Polarized distributions}
While the unpolarized collinear and TMD distributions can be extracted from global fits to data, the polarized Boer-Mulders' and Collins' functions arise in sub-leading effects making them harder to extract.
However, constraints on the positivity of the density matrix lead to a bound on the polarized distributions known as the Soffer bound \cite{Soffer:1994ww,Anselmino:2004ky}
\begin{align}
	h^{\bot (1)}_q(x,k_\bot^2) \equiv&
	\frac{k_\bot}{2M_p} h^{\bot}_q(x,k_\bot^2) \leq f_q(x,k_\bot^2)\;,\\
	h^{\bot (1)}_g(x,k_\bot^2) \equiv&
	\frac{k_\bot^2}{2M_p^2} h^{\bot}_g(x,k_\bot^2) \leq f_g(x,k_\bot^2)\;,\\
	H^{\bot (1)}_q(x,k_\bot^2) \equiv&
	\frac{k_\bot}{2M_p} H^{\bot}_q(x,k_\bot^2) \leq D_q(x,k_\bot^2)\;,\\
	H^{\bot (1)}_g(x,k_\bot^2) \equiv&
	\frac{k_\bot^2}{2M_p^2} H^{\bot}_g(x,k_\bot^2) \leq D_g(x,k_\bot^2)\;.
\end{align}

For the present work, we will consider the polarized distributions to be proportional to the unpolarized distributions, i.e. $h^{\bot (1)}(x,k_\bot^2) = b f_q(x,k_\bot^2)$ and $H^{\bot (1)}(x,k_\bot^2) = B D_q(x,k_\bot^2)$.
Since all the processes that we consider are proportional to the product of two polarized distributions, we will only modulate the overall factor of\footnote{For the process with two Boer-Mulders functions we will consider that $B\cdot B = B \cdot b$.} $B\cdot b$.
Once may assume that $B\simeq b$. We present results for different values ($0.2\leq B\cdot b \leq 0.4$) in Sec.~\ref{sec:Results}.

\section{Bulk simulations and elliptic flow}
\label{sec:simulations}

In the preceding section, we have described the 2-to-2 scattering of polarized and unpolarized partons ($a+b\rightarrow c+d$), with transverse momentum in the initial state. If the two out-going hard partons have transverse momenta  $\bm{p}_{T,c}$ and $\bm{p}_{T,d}$, then there \emph{will} be an asymmetry in the azimuthal angle $\phi$ between $\bm{q}_T = \bm{p}_{T,c} + \bm{p}_{T,d} $ and 
$\bm{Q}_T = \bm{p}_{T,c} - \bm{p}_{T,d} \approx 2 \bm{p}_{T,c} $~\cite{Boer:2010zf, Pisano:2013cya, Hatta:2020bgy}.

However, the angle $\phi_{q_T}$, which indicates the direction of the vector $\bm{q}_T$, varies from event to event. The azimuthal asymmetry in the angle between $\bm{q}_T$ and $\bm{Q}_T$ will yield anisotropy coefficients ($v_n$) if there is a correlation between $\phi_{q_T}$ and the event plane angles $\Psi_n$. That is, if the system can be reoriented in each event, based on the detected particles in that event. 
In this section, we will demonstrate that $\phi_{q_T}$ is strongly correlated with the event plane angles of the colliding system.

\subsection{Elliptic flow in small systems}
\begin{figure}
\includegraphics[width=0.45\textwidth]{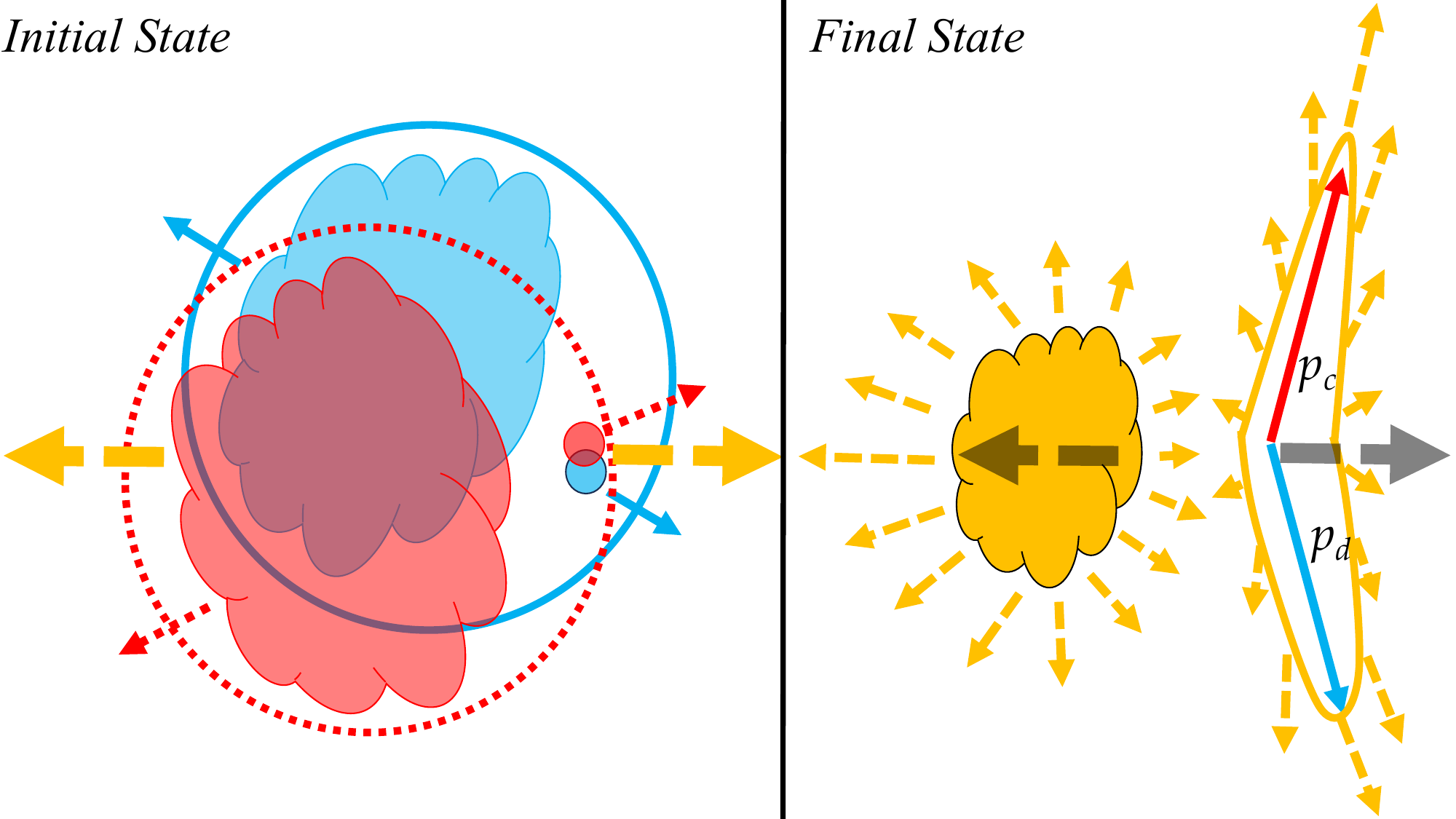}
\caption{ Left: A cartoon of a $p$-$p$ collision with relative transverse momentum in the initial state. The blue proton indicated by the solid blue circle comes out of the page, and contains a hard parton (blue dot) with transverse momentum, recoiling off the rest of the proton represented by the blue blob moving in the opposite direction. The net transverse momentum of the proton is zero, and so the hard parton and the remainder of the proton are balanced in transverse momentum. The red proton going into the page has a similar situation with its hard parton and remainder also balanced in transverse momentum. Right: the final state after the collision can be separated in a part containing the dijet pair resulting from the scattering of the two hard partons recoiling off the remainder of the colliding system. 
}\label{fig:v2-illustration}
\end{figure}

Azimuthal angular distributions of particles are expressed in terms of the Fourier series
\begin{align}\label{eq:FouierSeries}
\frac{dN}{d\phi dp_T dy} = \frac{dN}{dp_T dy} \left(1+ 2 \sum_n v_n \cos[n(\phi-\Psi_n)]\right) \;,
\end{align}
where the $\Psi_n$ are event plane angles, with $dN$ the differential yield of particles with momentum given as 
$p_z = \sqrt{p_T^2 + m^2} \sinh (y)$ and $p_x = p_T \cos(\phi) , p_y = p_T \sin(\phi)$.

In the case of a heavy-ion collision, the origin of azimuthal anisotropies in the momentum distribution of particles is now well understood: An initial geometric asymmetry, i.e. a spatially asymmetric density distribution leads to pressure gradients, which in turn lead to an asymmetric momentum distribution.
In the case of small systems, there is no geometrical asymmetry in the initial state. In events without hard scattering, any asymmetries are generated entirely from fluctuations in the initial state~\cite{Mantysaari:2017cni}. However, the presence of a hard scattering leads to a much larger multiplicity,  that overwhelms the yield of particles produced in entirely soft interactions~\cite{Buttar:2005gdq}. In the case of collisions with hard partonic scattering, multiplicity or transverse energy  production at almost all rapidities is dominated by hadrons that originate in the hard scattering, or are connected, in some way, to the hard scattering, e.g., in the breaking of strings that stretch from the hard high $p_T$ parton to the remnants that go down the beam line~\cite{Sjostrand:1987su}. 

Particle production, and the eventual momentum asymmetries, that are effected by the hard sub-process, may be described as follows: Let us consider a hard parton carrying momentum fraction $x_a$ of the proton $A$.
If the hard parton possesses, in addition, a transverse momentum $\bm{k}_{\bot, a}$, orthogonal to the proton's momentum, the remaining partons (and any other constituents) inside the proton must have the same magnitude of transverse momentum, but in the opposite direction, i.e., $-\bm{k}_{\bot, a}$.

On the left side of Fig.~\ref{fig:v2-illustration} we present the initial state of a proton-proton scattering where the blue proton with a solid boundary is coming out of the page, the red proton with the dotted boundary goes into the page.
The hard parton is indicated by the solid blue/red dot and the remainder of the proton is indicated by the blue/red blob. The balanced transverse momenta between the hard parton and the remainder, in the blue proton, are indicated by the opposing blue arrows. 
The opposing red arrows indicate a similar situation in the red proton. 
In the illustration, we show the hard parton as physically separated from the remainder of the photon. This is just for illustration, the hard parton may indeed be within the vicinity of the remainder.

After the scattering, hard partons produce 2 back-to-back jets in the right panel of Fig.~\ref{fig:v2-illustration}. 
The center of mass of the final state partons is shifted by the transverse momentum $\bm{q}_T=\bm{k}_{\bot, a} + \bm{k}_{\bot, b} = \bm{p}_{\bot, c} + \bm{p}_{\bot,d}$.
While the entire remainder does not participate in the collision, one expects that the total transverse momentum of the bulk is generally opposite to the hard scattering $\sim -\bm{q}_T$. 

In the illustration described above (and in Fig.~\ref{fig:v2-illustration}), we have considered the case of a single 2-to-2 hard scattering, with the remainder of the colliding system being soft. This does not have to be the case, the remainder of the participating system may engender multiple hard scatterings as well. In this case, the dots in our illustration indicate the two partons that undergo the hardest exchange. The blob will represent the remainder of hard and soft scatterings. None of our considerations above are affected by this. 
We will investigate these assumptions further by considering a simple PYTHIA~\cite{Bierlich:2022pfr} simulation in the subsequent subsec.~\ref{sec:PYTHIA}.

\subsection{PYTHIA simulations}\label{sec:PYTHIA}

In this paper, we have studied the azimuthal distribution of high-$p_T$ hadrons produced in the scattering of hard partons that possess transverse momentum in the initial state. To experimentally determine this azimuthal distribution, one needs a reference direction in the azimuthal plane. Since the incoming state consists of unpolarized protons, no reference angle is provided by the initial state; it has to be determined from the distribution of hadrons in the final state. 

In the preceding subsection, we argued that the vector $\bm{q}_T = \bm{k}_{\bot, a} + \bm{k}_{\bot, b} = \bm{p}_{\bot, c} + \bm{p}_{\bot,d}$ would be correlated with the event plane angle(s). In this subsection, we demonstrate this with a PYTHIA simulation~\cite{Bierlich:2022pfr}. As will be demonstrated in the next section, the largest component of our calculated azimuthal asymmetry is for the case of polarization dependent scattering. Such simulations are not possible using PYTHIA, or any other event generator that correlates hard and soft particle production. 
Our simulations will thus demonstrate the angular correlation only for the case of polarization averaged scattering, which will constitute a small component of the calculated azimuthal anisotropy. Also, in PYTHIA simulations with multiple interactions turned on, the initial transverse momentum of each parton that eventually encounters a hard scattering, is balanced of its sole remnant and not distributed among other partons and remnants in the same proton.

To overcome the above limitations of typical PYTHIA simulations, we set up events where initial state radiation, final state radiation and multiple parton interactions are turned off. Each proton now consists of a hard parton that will participate in the hard scattering (small dots in Fig.~\ref{fig:v2-illustration}), other partons that do not participate in the hard scatterings, and diquarks, that along with the non-participating partons, constitute the remnant or reminder of the proton (blobs in Fig.~\ref{fig:v2-illustration}).

The entire events thus consist of two hard partons which undergo a hard scattering and remnants that proceed down the beam line. 
To simulate TMD effects, we generate a transverse momentum $k_{Ti}$ for each remnant, from a Gaussian distribution with a width of $\langle k_\bot^2 \rangle^{1/2}$.
The total transverse momentum of all remnants $\bm{q}_T=\sum_i \bm{k}_{Ti}$ is compensated by applying a Lorentz boost to the center of mass frame of the hard scattering with the velocity
\begin{align}
	\bm{v} = \frac{\bm{q}_T}{\sqrt{(E_a + E_b)^2 + \bm{q}_T^2}}\;.
\end{align}
This leads to initial state partons with a transverse momentum following a Gaussian distribution with a width comparable to the remnants' width.
By using a boost to generate the transverse momentum of the initial state partons, we ensure that the kinematics of the scattering is unmodified, such that we ensure energy-momentum conservation in the hard-scattering while balancing the transverse momentum between the hard scattering and the remnants partons.

The above equation also does not work in the case there is initial state radiation, as the hard partons are then not directly balanced off the remnants. 
The remnants and final state partons are hadronized in PYTHIA using the Lund string model \cite{Andersson:1983ia}, where strings are connected between the hard outgoing partons and the beam remnants. 

\subsection{Event plane angle}

As described in the previous subsection, we have generated events where the hard scattering system of two partons has a net transverse momentum, $\bm{q}_T = \bm{p}_{\bot, c} + \bm{p}_{\bot, d}$, where $\bm{p}_{\bot, c} $ and $\bm{p}_{\bot, d}$ are the large momenta of the outgoing partons, which are now no longer back-to-back. The remainder of the event, which consists of beam remnants that are connected via strings with the hard partons and remnants which are not connected to a hard scattering, will recoil with a net transverse momentum $-\bm{q}_T$. 

\begin{figure}
	\begin{center}
		\includegraphics[width=0.22\textwidth]{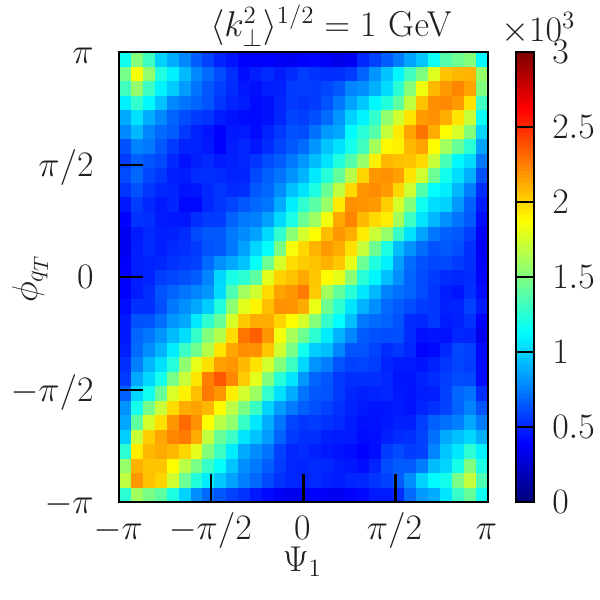}\includegraphics[width=0.22\textwidth]{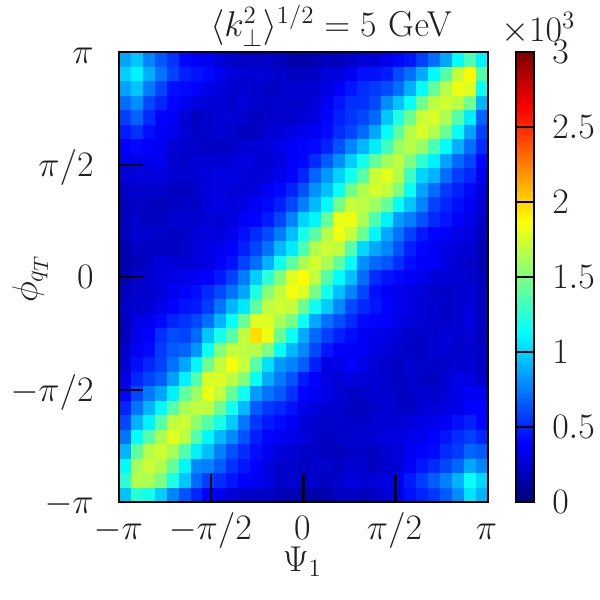}
		\includegraphics[width=0.22\textwidth]{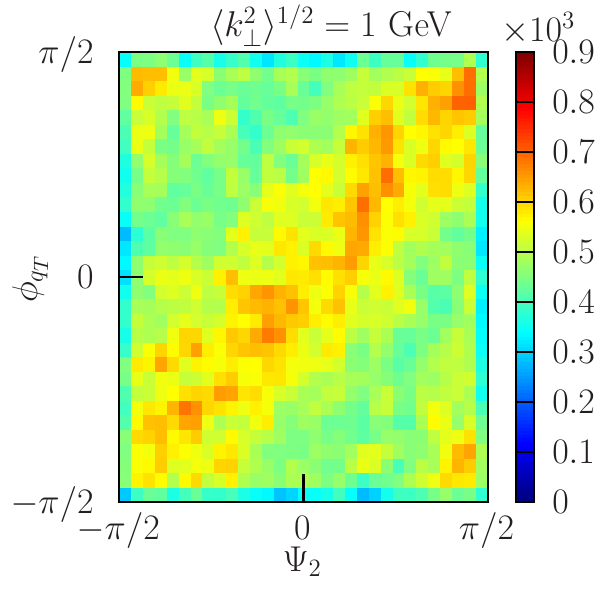}\includegraphics[width=0.22\textwidth]{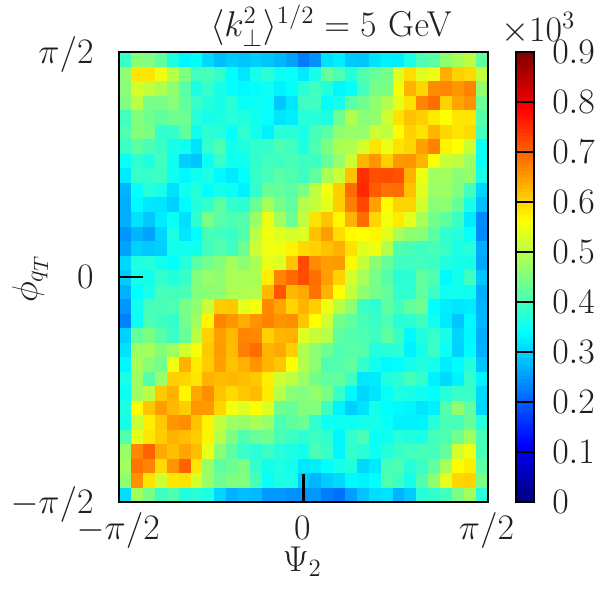}
	\end{center}
	\caption{
		Histogram of the correlation between the event plane angle and the azimuthal angle of the total transverse momentum for $\langle k_\bot^2\rangle^{1/2} = 1$ GeV (left) and $\langle k_\bot^2\rangle^{1/2} = 5$ GeV (right) using the event plane angle $\Psi_1$ (top) and $\Psi_2$ (bottom).
	}\label{fig:EventPlane}
\end{figure}

Once the entire event is hadronized, 
the event plane angle $\Psi_n$ in Eq.~(\ref{eq:FouierSeries}) can be obtained using the flow vector $\bm{Q}_n$ \cite{Borghini:2001vi} defined as follows,
\begin{align}
	\bm{Q}_n 
	= \frac{1}{\sqrt{M}} \sum_j^{M} e^{in\phi_j}\;,
\end{align}
where the sum runs over all hadrons in the event (typically over some rapidity range) and $\phi_j$ is the azimuthal angle of the hadron $j$.
The event plane angle is then defined as the Euler angle of the flow vector $\bm{Q}_n$ as follows
\begin{align}
	\Psi_n 
	= \frac{1}{n} \arg [\bm{Q}_n]\;.
\end{align}

In Fig.~\ref{fig:EventPlane}, we compute correlation histograms of the values of the event plane angle $\Psi_n$ and the azimuthal angle of the total transverse momentum $\bm{q}_T$.
For both $\Psi_1$ and $\Psi_2$, we observe a clear correlation between the two angles which becomes more pronounced as the mean transverse momentum is increased $\langle k_\bot^2\rangle^{1/2} = 5$ GeV.
This implies that in the presence of hard scattering, leading to two high $p_T$ jets, the net transverse momentum of the jets $\bm{q}_T = \bm{p}_{\bot, c} + \bm{p}_{\bot, d}$, is tightly correlated with the $1^{\rm st}$ and $2^{\rm nd}$ order event plane (it is also correlated with higher order event planes, though more weakly). Thus, in events with a hard scattering, the direction of the $\bm{q}_T$ vector can be determined, by analyzing all the hadrons in an event, and thus the azimuthal distribution of $\bm{p}_{\bot c/d}$ with respect of $\bm{q}_T$ can indeed be determined.

\subsection{Centrality selection}

While not a central issue in our calculations, our PYTHIA simulations reveal another interesting feature of these small system collisions: The correlation between the transverse momentum of the incoming partons and the transverse energy produced in the rapidity of forward calorimeters.  
There are several methods used by experiments to select the centrality of the collision.
One such method is based on the transverse energy in the forward calorimeter (FoCal) which measures hadrons in the pseudo-rapidity range $3.9\leq|\eta|\leq 5.0$.
Transverse momentum of the initial hard partons can lead to additional transverse energy in the FoCal.

\begin{figure}[!h]
    \centering
    \includegraphics[width=0.45\textwidth]{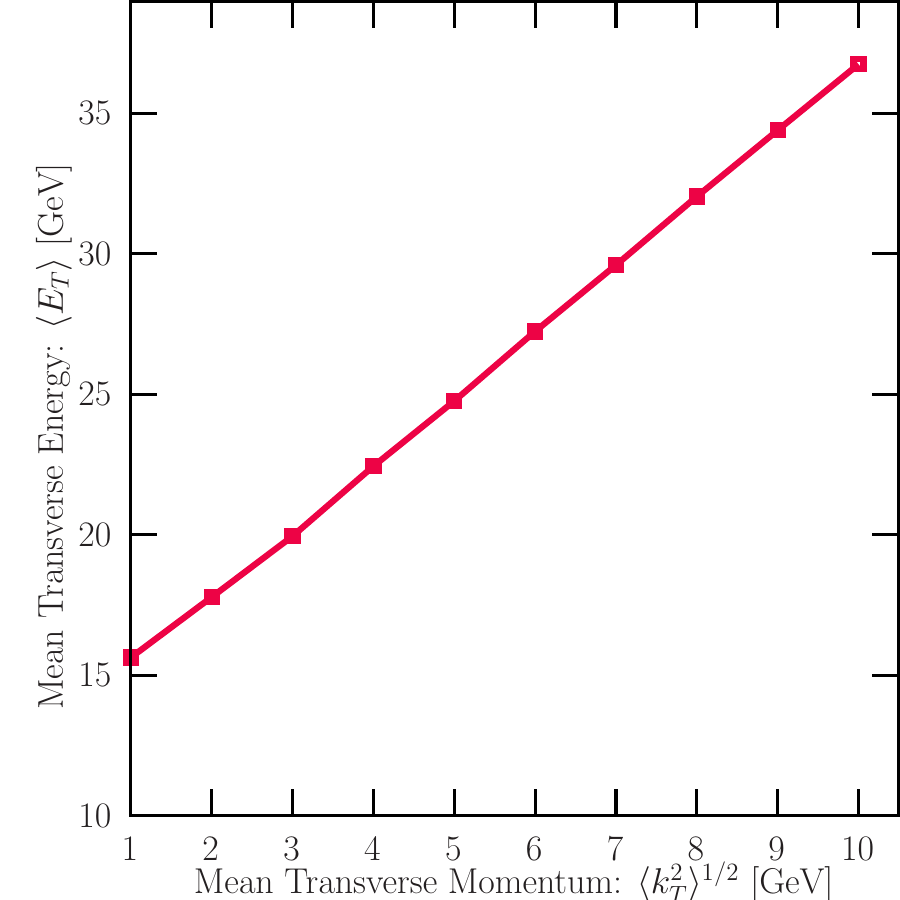}
    \caption{Average transverse energy $\langle E_T \rangle = E \sin\theta$ in the forward direction $3\leq|\eta|\leq 10$ as a function of the average transverse momentum of the initial hard parton $\langle k_T^2\rangle^{1/2}$. }
    \label{fig:AverageET}
\end{figure}

Using the results of the simulations in the previous subsection, we compute the transverse energy in each event as, 
\begin{align}
	E_T = \sum_i E_i \sin\theta_i\;,
\end{align}
where $E_i$ is the energy of the final state hadron and $\theta_i$ is the angle between the hadron momentum and the $z$-axis. The sum runs over all hadrons in the event within a prescribed range in (pseudo-)rapidity. 
The average transverse energy $\langle E_T \rangle$ is shown in Fig.~\ref{fig:AverageET} as a function of the average transverse momentum of the initial hard parton $\langle k_T^2\rangle^{1/2}$.
We find a clear linear relation between the initial transverse momentum of the hard parton and the transverse energy in the forward direction.

Experimental data on the elliptic anisotropy, from the ATLAS detector~\cite{ATLAS:2019vcm}, that will be compared against, are extracted from the most central (0-5\%) events in $p$-$Pb$ collisions. In these experiments, centrality is determined from the amount of transverse energy deposited in the FoCal. The results of our simulations above indicate that larger transverse energy deposited in the FoCal isolates events where the mean intrinsic transverse momentum of the initial state partons is larger. These are events where the correlation between the direction of $\bm{q}_T$ and the event plane angles is more enhanced (See Fig. \ref{fig:EventPlane}). This further enhances our argument that the angle of $\bm{q}_T$ can be determined in experiment.

\section{Results}\label{sec:Results}

In the previous section, we argued that events with hard scattering that contain a dijet pair recoiling off the remainder of the colliding protons will have a well defined event plane, with several event plane angles coinciding with the angle of the dijet momentum $\bm{q}_T = \bm{p}_{\bot, c} + \bm{p}_{\bot, d}$ (where $\bm{p}_{\bot, c}$ and $\bm{p}_{\bot, d}$ are the large transverse momenta of the out-going jets, see Fig.~\ref{fig:v2-illustration}). We further argued that more central events isolated via larger transverse energy deposition in the forward calorimeter, enhances events with larger intrinsic transverse momentum in the initial state, leading to events with a larger $\bm{q}_T$. 

\begin{figure}
\centering
\includegraphics[width=0.45\textwidth]{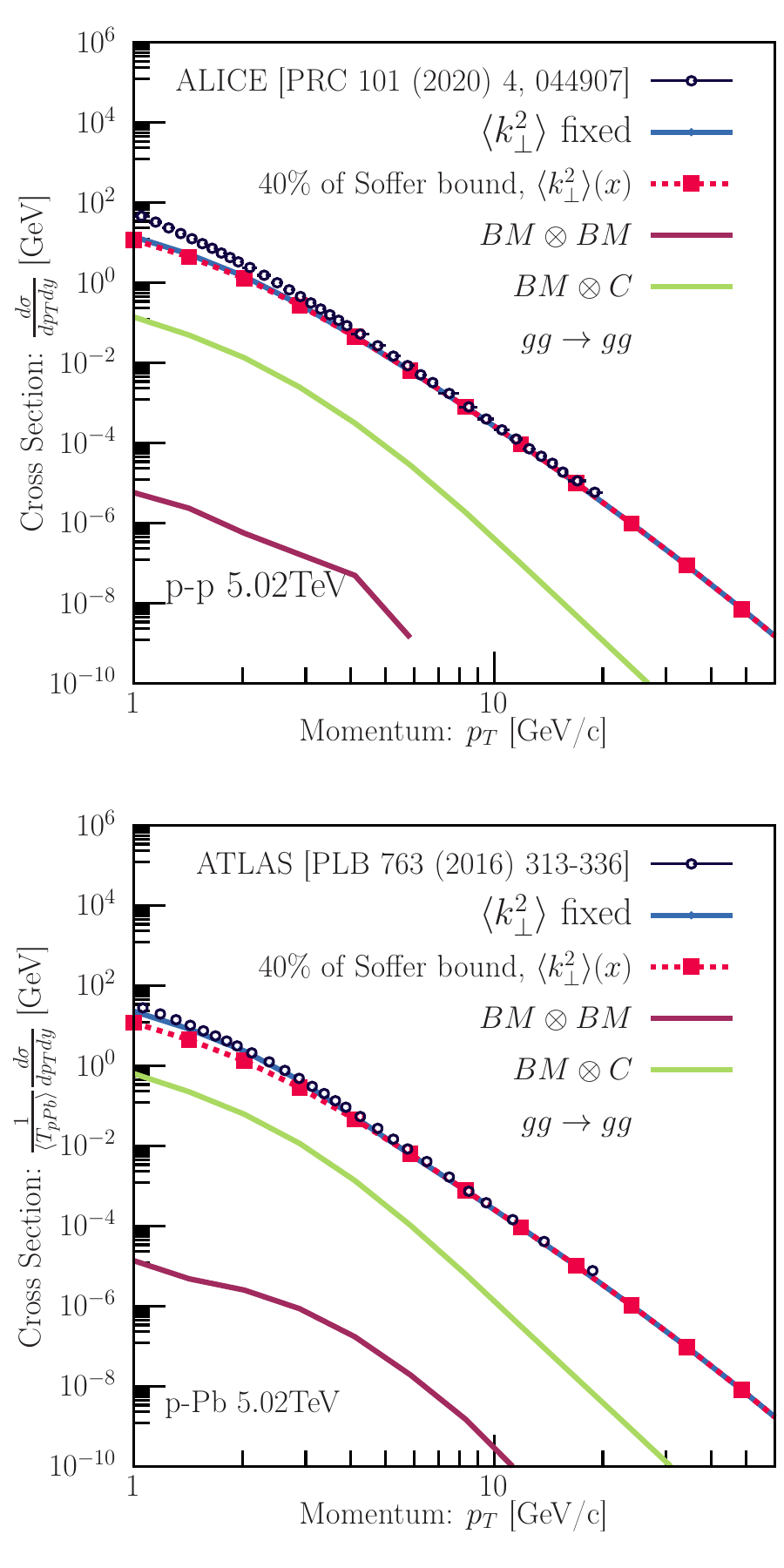}
\caption{ 
	Pion production cross section using gluon-gluon scattering only in $p$-$p$ at $5.02$ TeV (top) and $p$-$Pb$ at $8.16$ TeV (bottom).
	The unpolarized contribution (red dashed line squares) is dominant over the $BM \otimes C$ (purple lines) and the $BM \otimes BM$ contributions (green lines).
	The cross section using an $x$-dependent $\langle k_\bot^2\rangle$ reproduces the collinear results (blue lines) at high $p_T\leq 5$ GeV.
	Experiment results from ALICE~\cite{ALICE:2019hno} and ATLAS~\cite{ATLAS:2016xpn}.
}
\label{fig:PionProduction}
\end{figure}

In the calculations that follow, we will assume that $\bm{q}_T$ \emph{is coincident} with the event plane angle(s). In reality, the event plane angle will be strongly correlated with the direction of $\bm{q}_T$. 
To compensate for this assumed coincidence, we will keep the mean $\langle k_\bot^2 \rangle$, for the case of the $x$-dependent $k_\bot$ distribution, somewhat below $1$~GeV in the case of $p$-$p$ collisions, as shown in Fig.~\ref{fig:MeankT}. This $x$-dependent $k_\perp$ distribution is used for the preponderance of spectra and azimuthal anisotropy calculations that follow, and is enhanced by a factor of $A^{1/3}$ for the case of $p$-$Pb$. For comparison purposes only, we also present the pion spectrum in $p$-$p$ and $p$-$Pb$ collisions (Fig.~\ref{fig:PionProduction}) using the fixed $\langle k_\perp^2 \rangle = 1$~GeV$^2$   prescription.
Thus, we are using values of $\langle k_\bot^2 \rangle$ that are closer to the lower limit considered in our PYTHIA simulations. 

Beyond this, there will be no further reference to the PYTHIA simulations. No further simulations will be carried out. Our results will be from analytical calculations. The anisotropy coefficients will be calculated assuming that $\bm{q}_T$ coincides with the direction of the event plane. We remind the reader that a majority of the calculated anisotropy will result from polarization dependent scattering; and currently, there is no event generator that can simulate polarization dependent processes. Thus, there is no alternative other than semi-analytical calculations.

\begin{figure}
\includegraphics[width=0.45\textwidth]{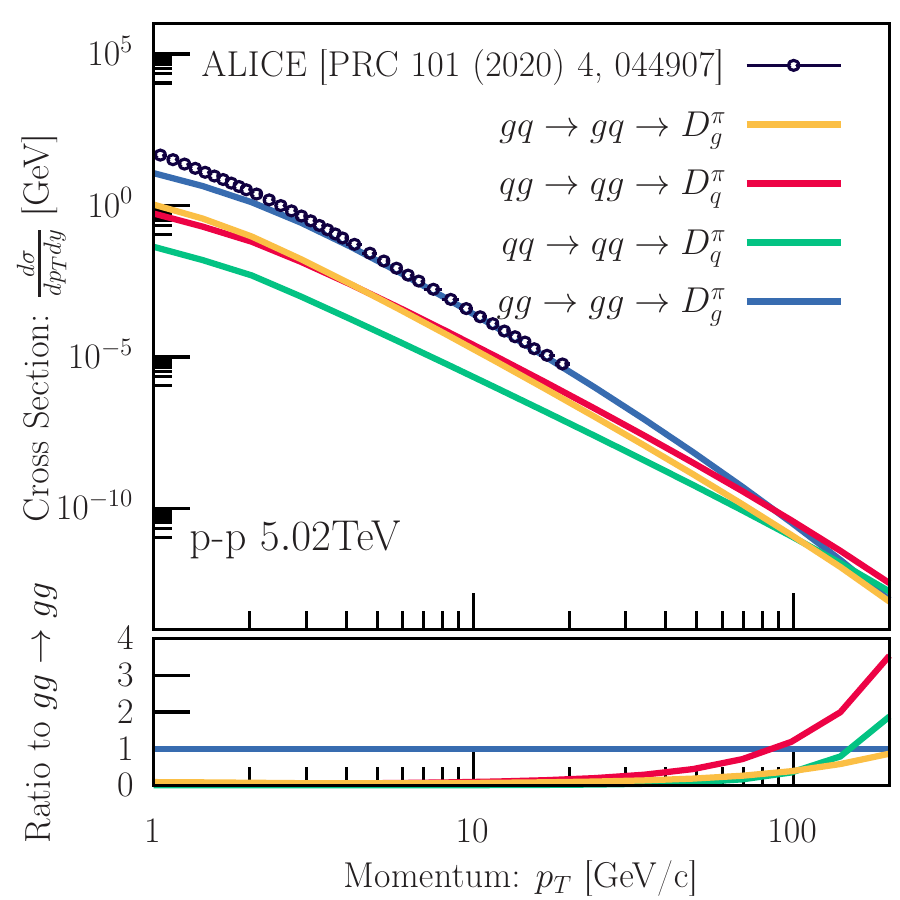}
\caption{
	Pion production cross section in $p$-$p$ at $5.02$ TeV.
	The top panel displays unpolarized contributions from $gg\to gg$ (blue), $qg\to qg$ (red), $qq\to qq$ (green) and $gq\to gq$ (orange) partonic scatterings.
	The bottom panel displays the ratio to the $gg\to gg$ scattering which dominates the cross section for $p_T\lesssim 90$ GeV.
	Compared with ALICE data \cite{ALICE:2019hno}.
}\label{fig:CrossSection}
\end{figure}

\begin{figure}
	\begin{center}
		\includegraphics[width=0.45\textwidth]{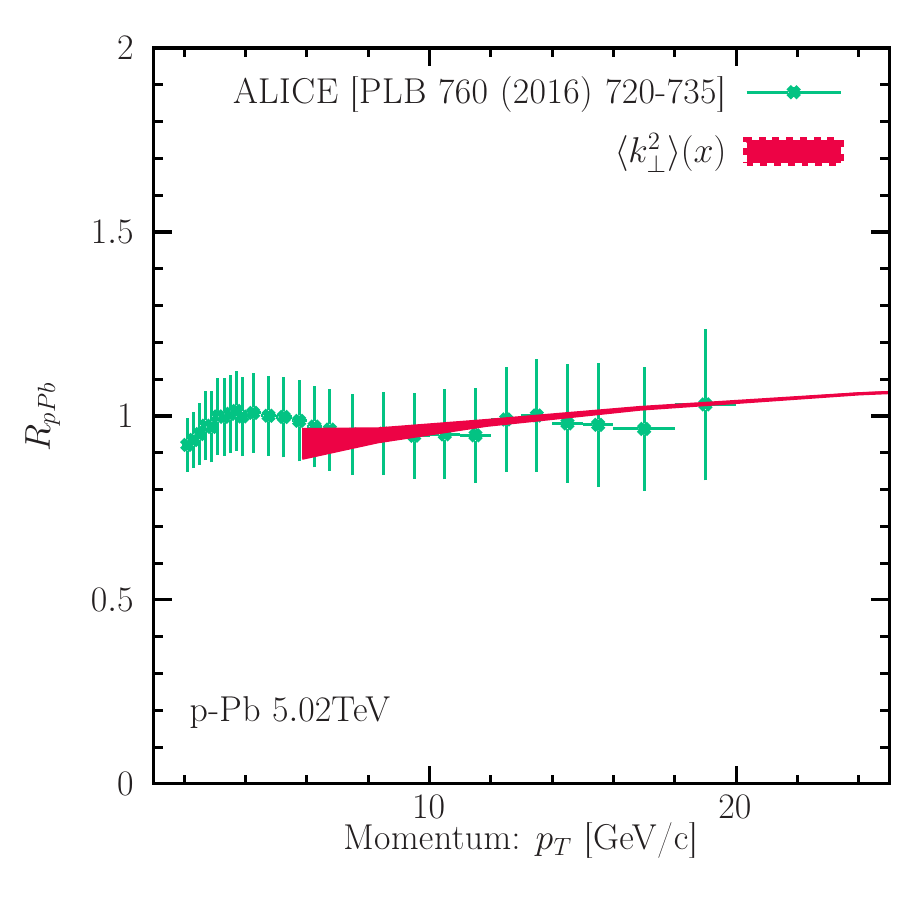}
	\end{center}
	\caption{Nuclear modification factor $R_{pPb}$ for pion production in $p$-$Pb$ at $5.02$~TeV.
		We compare with ALICE results~\cite{ALICE:2016dei}.
	}
	\label{fig:RpPb}
\end{figure}

For our calculations, we employ the nCTEQ parameterization~\cite{Kovarik:2015cma} for integrated PDFs [$f(x,Q^2)$] and leading order KKP~\cite{Kniehl:2000fe} for FFs.
While we will use either a constant or $x$-dependent mean transverse momentum for the TMD-PDFs, we will only employ a constant $\langle k_\bot^2\rangle^{1/2}=0.8$ GeV for the pion's TMD-FFs, following the work of Ref.~\cite{DAlesio:2004eso,Anselmino:2005sh}.

When using a constant transverse momentum $\langle k_\bot^2\rangle$ the Gaussian $k_\bot$ dependence integrates out, recovering the results from collinear factorization.
We display the angular integrated cross section, using full lines in Fig.~\ref{fig:CrossSection}, for different partonic scatterings $ab\to cd$, with the fragmenting parton $c$.
As expected, the gluon PDFs are dominant at lower values of the $Q^2$ scales; this is partially responsible for the $gg\to gg$ partonic scattering contribution remaining dominant up to $p_T\lesssim 80$ GeV, beyond which quark initiated processes start to contribute. We also include the experimental data from the ALICE collaboration~\cite{ALICE:2019hno}, which, given the limited range in $p_T$, is almost completely described by the $gg \to gg$ process.

In Fig.~\ref{fig:PionProduction}, we further decompose the $gg \to gg$ cross section with $x$-dependent 
$\langle k_\perp^2 \rangle $ into the polarization averaged (dashed red line squares), and polarization dependent contributions which include a Boer-Mulders distribution with a Collins fragmentation function ($BM\otimes C$ in green solid) and a double Boer-Mulders distribution ($BM \otimes BM$ in maroon solid). The upper panel represents $p$-$p$ at $\sqrt{s_{\rm NN}} = 5.02$~TeV, while the lower panel is for $p$-$Pb$ collisions, at $\sqrt{s_{\rm NN}} = 5.02$~TeV.
In blue lines we present the cross section for fixed $\langle k_\perp^2\rangle$ which recovered the collinear result as the Gaussian profile integrates to unity.

Since polarization phases do not cancel for the polarized cross sections, the angular integrated spectrum is suppressed compared to the unpolarized cross section.
The $x$-dependence leads to a deviation from the collinear cross section at low $p_T$, visibly for the case of $p$-$Pb$ (and barely for the case of $p$-$p$). 
However, at $p_T \gtrsim5$ GeV, the effect of the transverse momentum on the cross section becomes negligible.
Above this $p_T \; (\gtrsim 5$~GeV$)$, the cross section is indistinguishable from the collinear result as shown in Fig.~\ref{fig:PionProduction}. As mentioned in Sec.~\ref{Pheno-v2}, for the case of $p$-$Pb$, shown in the bottom panel of Fig.~\ref{fig:PionProduction}, the $\langle k_\perp^2 \rangle$ for the proton is enhanced by $A^{1/3}$ to account for multiple soft interaction in the $Pb$ prior to the hard scattering.

We present the nuclear modification factor $R_{pPb}$ in Fig.~\ref{fig:RpPb} for pion production.
No significant modification is observed between the $p$-$p$ and $p$-$Pb$ spectra consistent with the minimum bias ALICE experimental result~\cite{ALICE:2016dei}.

\begin{figure}
\begin{center}
	\includegraphics[width=0.45\textwidth]{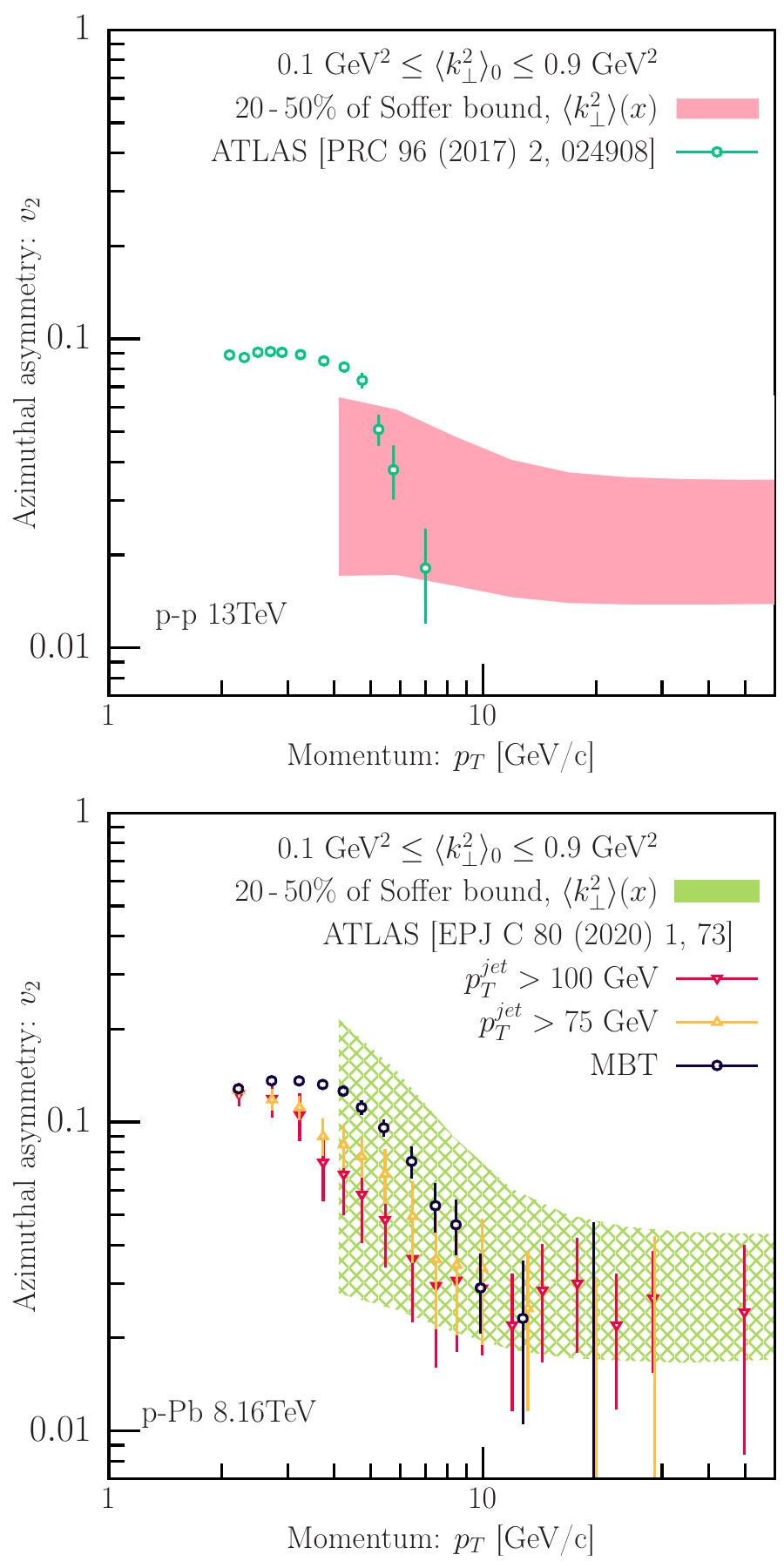}
\end{center}
\caption{
	Elliptic flow coefficient $v_2$ for pion production in $p$-$p$ at $5.02$ TeV (top) and $p$-$Pb$ at $8.16$ TeV (bottom).
	The green and red filled areas are the uncertainty over the functional form of the mean transverse momentum between $\langle k_\perp^2\rangle_0 = 0.1 - 0.9$~GeV$^2$ using $B\cdot b = 0.2-0.5$ for the polarized contributions.
}
\label{fig:v2pp}
\end{figure}

\begin{figure}
	\includegraphics[width=0.45\textwidth]{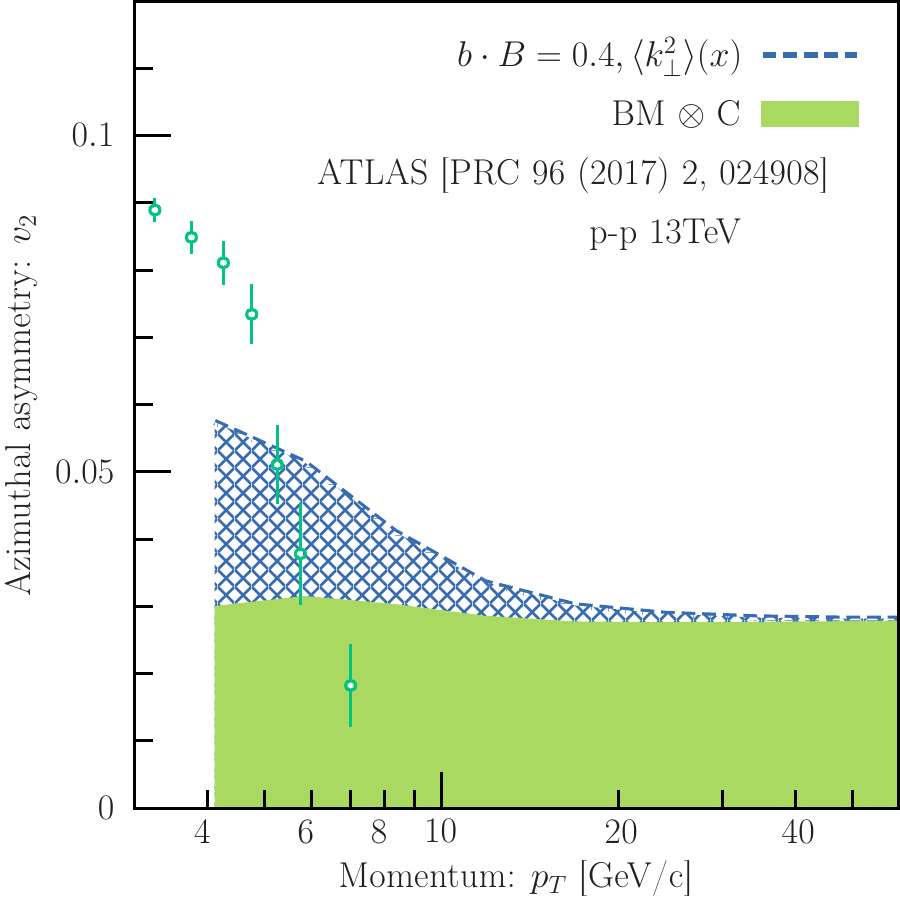}
	\includegraphics[width=0.45\textwidth]{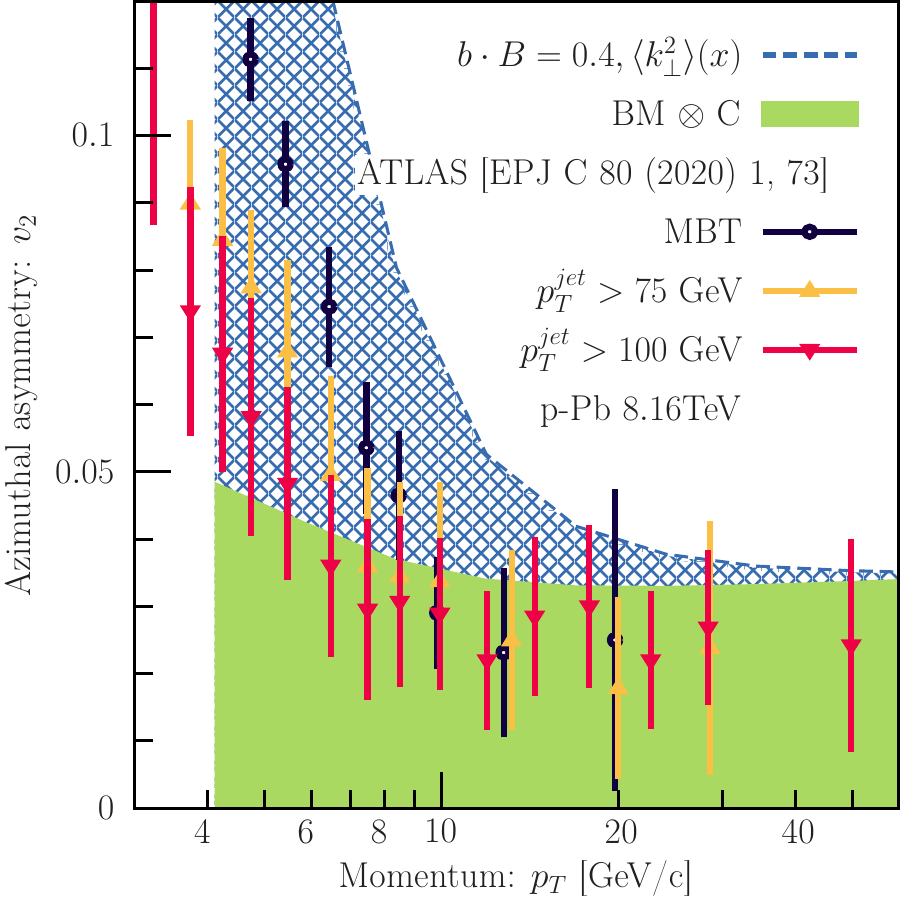}
	\caption{
		The different contributions to the elliptic flow coefficient $v_2$ in $p$-$Pb$ at $8.16$ TeV using the $x$-dependent $\langle k^2 \rangle^{1/2}(x)$ GeV and $B\cdot b = 0.4$ for the unpolarized distribution in $p$-$p$ (top) and $p$-$Pb$ (bottom).
		The hatched blue area is the unpolarized contribution, the green area is the $BM\otimes C$ terms.
	}\label{fig:v2Log}
\end{figure}

The cross section can be written as an expansion in the azimuthal angle $\phi_{\pi}$ as follows
\begin{align}
	\frac{d\sigma}{dp_T dy d\Delta\phi}
	=& \frac{d\sigma}{dp_T dy} \left(1+2\sum_{n=1} v_n \cos (n\Delta \phi)\right)\;,
\end{align}
where $\Delta \phi\equiv \phi_{\pi} - \phi_{\bm{q}_T}$ is the azimuthal angle of the pion $\phi_{\pi}$ with respect to the total transverse momentum of the hard scattering $\bm{q}_T$.
The flow coefficients $v_n$, can be obtained by integrating over the azimuthal angle $\Delta \phi$ as follows
\begin{align}
	v_n
	= \langle \cos(n\Delta \phi) \rangle
	=& \frac{\int d\Delta \phi~ cos(n\Delta \phi) \frac{d\sigma}{dp_T dy d\Delta\phi}}{\int d\Delta\phi~ \frac{d\sigma}{dp_T dy d\Delta\phi}}\;.
\end{align}
The elliptic flow coefficient $v_2$ obtained is displayed in Fig.~\ref{fig:v2pp} ($p$-$p$ in the top panel and $p$-$Pb$ on the bottom panel).
The uncertainty over the functional form of the mean transverse momentum is reflected in the red ($p$-$p$) and green ($p$-$Pb$) filled areas which is the bound between the mean transverse momentum $0.1$~GeV$\leq \langle k_\perp^2 \rangle_0 \leq 0.9$~GeV$^2$ and using $B\cdot b = 0.2-0.5$ for the polarized contribution. 
Within uncertainties, the elliptic flow coefficient $v_2$ is consistent with the experimental data from ATLAS \cite{ATLAS:2016yzd,ATLAS:2019vcm} for $p_T \gtrsim 6$~GeV.

To investigate the importance of the polarization dependent terms versus polarization averaged terms, we present the different contributions to the elliptic flow the $x$-dependent mean transverse momentum in Fig.~\ref{fig:v2Log} with $\langle k_\perp^2 \rangle_0 = 0.9$~GeV$^2$.
The unpolarized contribution is shown in blue hatched area, and the $BM\otimes C$ terms are shown in green area.
Although we use $B\cdot b = 0.5$ for the polarized contribution the $BM\otimes C$ terms are dominant compared to the unpolarized contribution at high-$p_T$. 
We observe that while the unpolarized contribution becomes negligible when $p_T \gg k_\perp $, the $BM \otimes C$ terms continue to contribute, and lead to a large azimuthal anisotropy coefficient, even at large transverse momentum.
This because the unpolarized contribution is power suppressed.

To further explore the azimuthal anisotropies arising from TMDPDFs, we compute higher order harmonics $v_3$ and $v_4$ which are displayed in Fig.~\ref{fig:v34pPb}.
The experimental results of $v_3$ and $v_4$ display a different behavior than $v_2$.
While $v_2$ starts decreasing with $p_T$ already at $p_T\gtrsim 5$ GeV, $v_3$ and $v_4$ are increasing up to the highest data points ($p_T\simeq 10$ and $p_T \simeq 4$, respectively).
However, the data displays large uncertainties at high-$p_T$ and is limited to $p_T \leq 10$ GeV.
While experimental data is only available for $p_T \leq 10$ GeV, the polarized contributions lead to non-zero coefficients at high-$p_T$.
Markedly, $v_4$ is entirely driven by the $BM \otimes BM$ terms at high-$p_T$, which does not contribute to any $v_n$ coefficient with $n<4$.

\begin{figure}
	\includegraphics[width=0.45\textwidth]{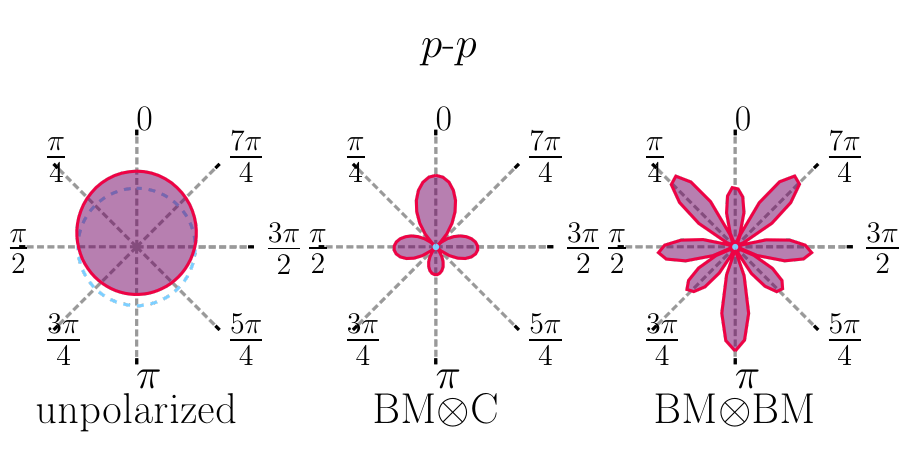}
	\includegraphics[width=0.45\textwidth]{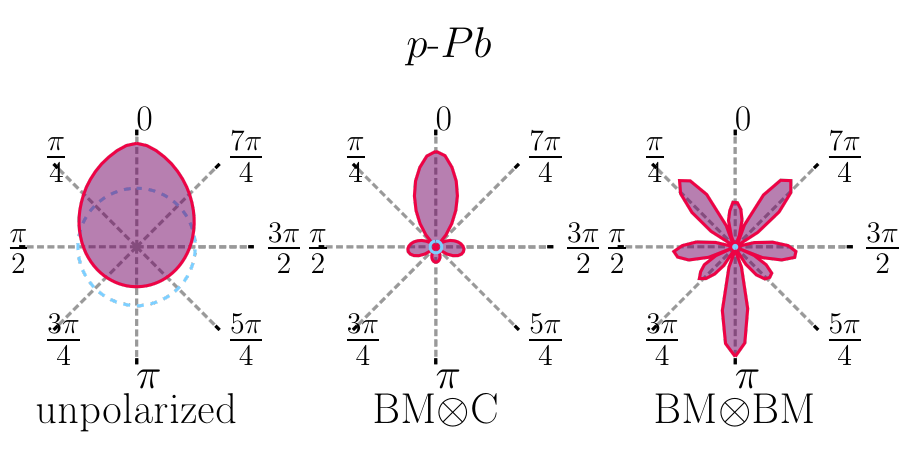}
	\caption{
		Azimuthal angle distribution $\frac{d\sigma}{d\phi}$ for $gg\to gg$ scattering pion production in $p$-$p$ at $5.02$ TeV (top) and $p$-$Pb$ at $8.16$ TeV (bottom) at a fixed $p_T=6$~GeV.
		The panels from left to right represents the unpolarized, $BM \otimes C$ and $BM \otimes BM$ contributions.
	}\label{fig:PhiDist}
\end{figure}
	
To obtain further insight into the origin of different $v_n$'s from different terms, we compute the azimuthal angular distribution of different contributions to the cross section, at a fixed $p_T$.
In Fig.~\ref{fig:PhiDist}, we present the azimuthal angle distribution $\frac{d\sigma}{d\phi}$ for $gg\to gg$ scattering in $p$-$p$ and $p$-$Pb$ collisions.
We find that for the $p$-$p$ scattering the unpolarized contribution has a large constant term and a small $v_1$ contribution.
While the $BM\otimes C$ and $BM\otimes BM$ terms contribute predominantly to $v_2$ and $v_4$, respectively.
Conversely, for $p$-$Pb$ collisions, while different terms display similar symmetries, they gain an additional elliptic shape due to the asymmetry of the transverse momentum distribution, leading to a larger $v_2$. We remind the reader that all terms in Fig.~\ref{fig:PhiDist} are for the process $gg \to gg$ (quark scattering contributions are included in the appendix).

\subsection{Predictions at RHIC}
\begin{figure}
	\begin{center}
		\includegraphics[width=0.45\textwidth]{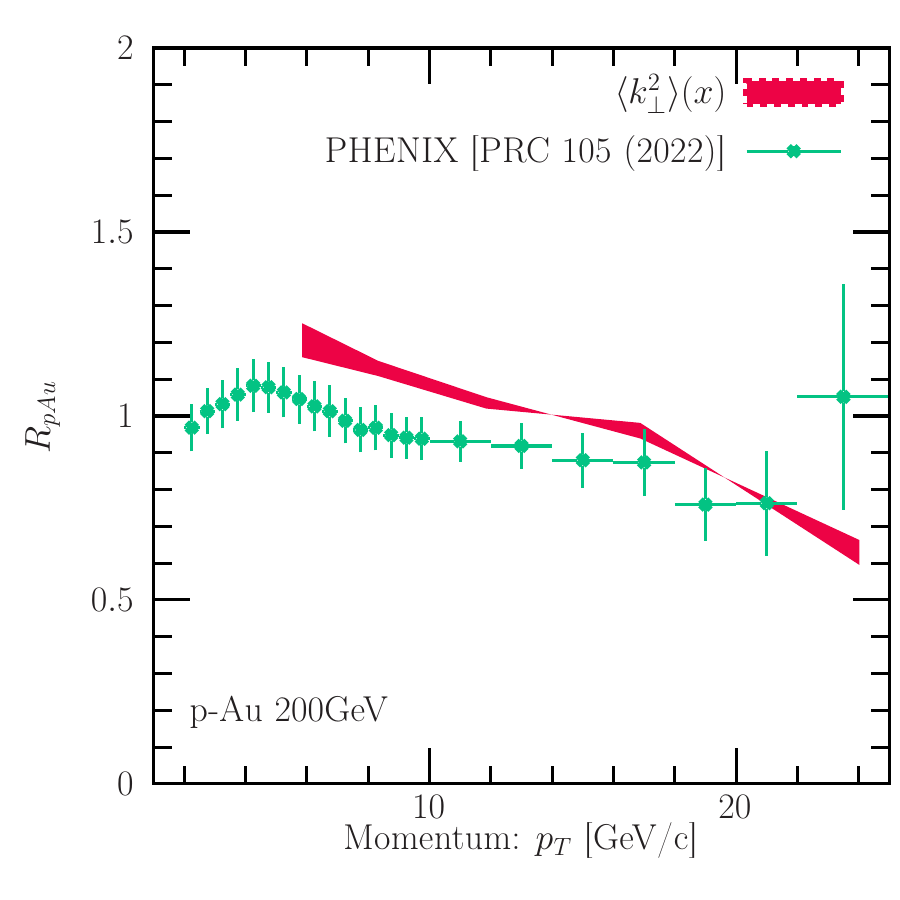}
	\end{center}
	\caption{Same as Fig.~\ref{fig:RpPb} for $p$-$p$ and $p$-$Au$ at $\sqrt{s}=200$~GeV. Calculations include shadowing as included in the nCTEQ PDF~\cite{Kovarik:2015cma}.
	The experimental points are from PHENIX~\cite{PHENIX:2021dod}.
	}\label{fig:RpAu}
\end{figure}
\begin{figure}
	\begin{center}
		\includegraphics[width=0.45\textwidth]{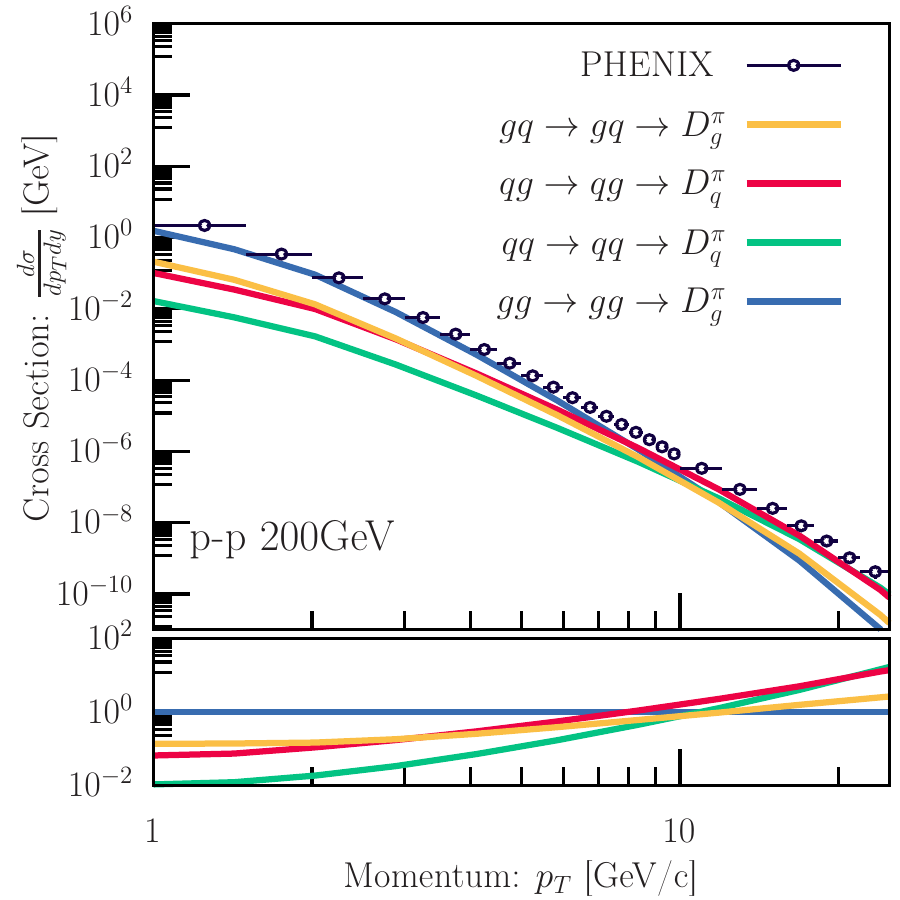}
	\end{center}
	\caption{Same as Fig.~\ref{fig:CrossSection} for $p$-$p$ at $\sqrt{s}=200$~GeV.}\label{fig:CrossSectionRHIC}
\end{figure}

\begin{figure}
	\begin{center}
		\includegraphics[width=0.45\textwidth]{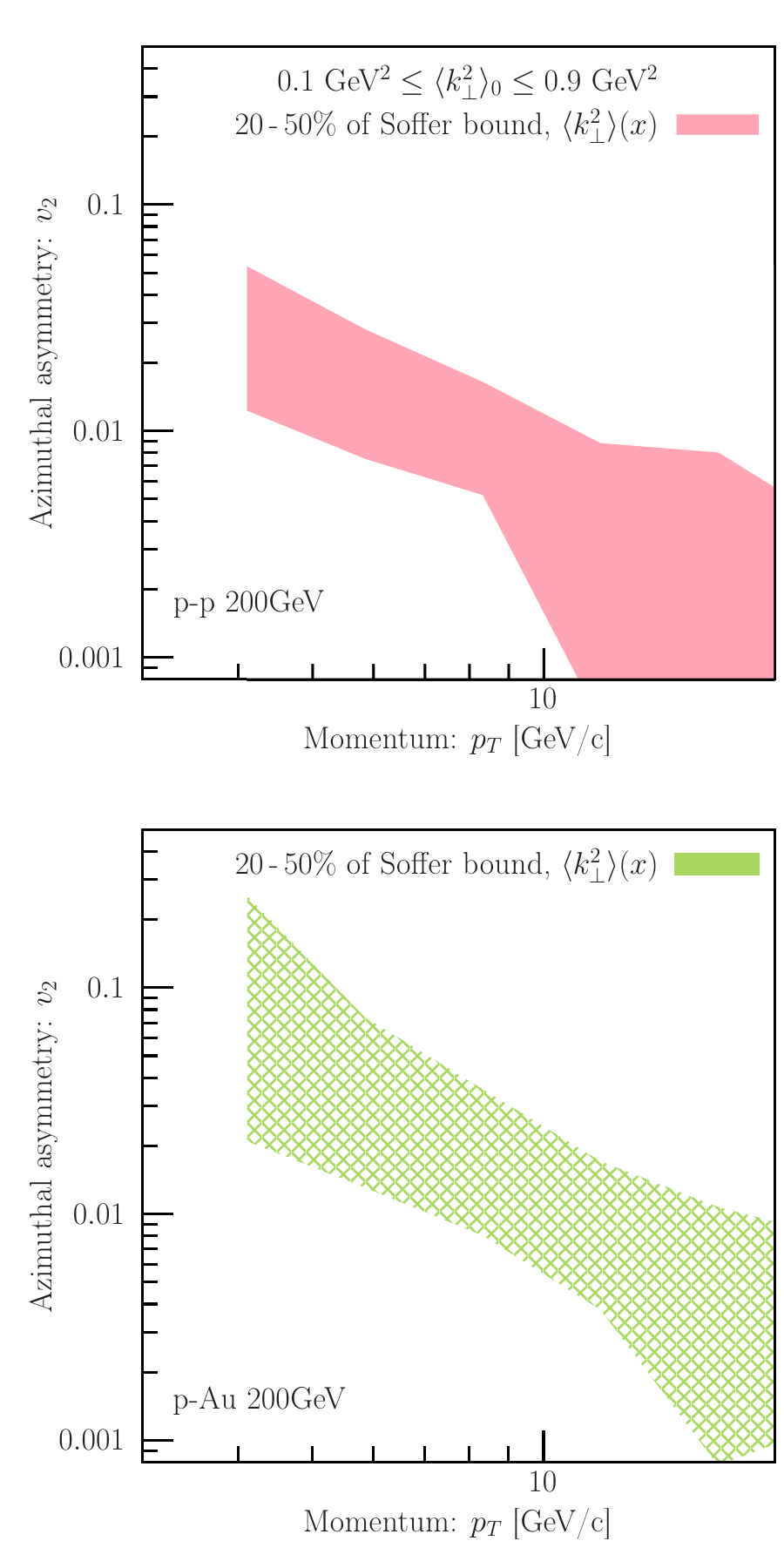}
	\end{center}
	\caption{Same as Fig.~\ref{fig:v2pp} for $p$-$p$ and $p$-$Au$ at $200$~GeV.}\label{fig:v2RHIC}
\end{figure}
We investigate azimuthal anisotropies at RHIC energies for $p$-$p$ and $p$-$Au$ collisions at $\sqrt{s_{\rm NN}} = 200$~GeV.
The nuclear modification factor $R_{pAu}$ was measured by the PHENIX collaboration~\cite{PHENIX:2021dod} and is shown in Fig.~\ref{fig:RpAu}.
While our results is restricted to LO matrix elements, they are compatible with NLO calculations obtained using the same nCTEQ PDFs in~\cite{PHENIX:2021dod}.
We observe minimal modification between $p$-$p$ and $p$-$Au$ collisions, consistent with the experimental data.

At RHIC energies with $\sqrt{s}=200$~GeV, the quark FFs become dominant at higher $p_T$ such that the quark initiated processes are more important as shown in Fig.~\ref{fig:CrossSectionRHIC}.
As such, we compute the azimuthal anisotropy $v_2$ for $p$-$p$ and $p$-$Au$ collisions at $\sqrt{s}=200$~GeV in Fig.~\ref{fig:v2RHIC} using both $gg \to gg$ and $qq \to qq$.
While the cross section for compton scattering is also important, the spin-dependent processes will require a BM$\otimes$C contribution on the quark line which should contribute a factor $\cos\phi$ which is less important for $v_2$.
We neglect these terms for our calculations where their contribution should be within our theoretical uncertainties.

The results are similar to what we observe at LHC energies, with non-zero $v_2$ at high-$p_T$.
However, limited by the collision energy both the cross section and the azimuthal anisotropy dies off at very high-$p_T$.
These results show that azimuthal anisotropies present a crucial avenue to explore the transverse momentum distribution of partons in the initial state at RHIC energies.

\section{Summary and Outlook}\label{sec:Conclusion}

In hard scattering processes calculated within the ambit of collinear factorization, the incoming partons possess no transverse momentum with respect to the direction of the incoming nucleons (in $p$-$p$ or $p$-$Pb$ collisions). In such processes, the outgoing partons and hadrons demonstrate no azimuthal anisotropy. However, in the event the incoming partons have a small amount of transverse momentum, anisotropies in the azimuthal distribution arise. The presence of transverse momentum in the initial state parton distribution, both with and without the presence of transverse polarization (linear polarization for gluons), leads to a scattering cross section that is not isotropic in azimuthal angle. 

In this paper, we have presented detailed results on the first attempt to calculate these anisotropies and compare them to experimentally measured Fourier coefficients (of these anisotropies) in $p$-$p$ and $p$-$Pb$ collisions. This attempt presents complete details of our calculations, which were briefly presented in Ref.~\cite{Soudi:2023epi}. In the range of $p_T$ explored by experimental data, the dominant channel is $gg \to gg$ (Fig.~\ref{fig:CrossSection}). Within this channel, the dominant contribution to the angle integrated cross section is the unpolarized process (Fig.~\ref{fig:PionProduction}). However, $gg \to gg$, with one gluon linearly polarized (Boer-Mulders' distribution), followed by the fragmentation of the linearly polarized gluon to an unpolarized hadron (Collins' function), provides the largest contribution to the elliptic coefficient $v_2$ (Fig.~\ref{fig:v2pp}), followed by the polarization averaged process (Fig.~\ref{fig:v2Log}). 

We calculate using a modified version of the $x$-dependent mean $\langle k_\perp^2 \rangle$ model of TMDPDF from Ref.~\cite{DAlesio:2004eso}.
The typical mean $\langle k_\perp^2 \rangle$ is set to be $\sim 0.9$~GeV$^2$ in $p$-$p$ for the x-dependent \emph{ansatz} (Fig.~\ref{fig:MeankT}).
These values are enhanced by $A^{1/3}$ of the $p$ in $p$-$A$ to account for pre-scattering off multiple nucleons prior to the hard exchange that produces the dijet. 

The $A^{1/3}$ enhancement required a factorized form for the TMDPDF (separate functions for $x$ and $k_\perp$ dependence), as multiple scattering based calculations for polarization dependent TMDs do not exist at this time. We further vary the ratio between the polarization dependent and independent distributions ($0.2 \leq b \cdot B \leq 0.5 $) to obtain a range of results for $v_2$. This is done as the exact magnitudes of the Boer-Mulders and Collins functions are not well known, at this time. In the absence of a bulk simulation which determines the event plane, we have argued in Sec.~\ref{Pheno-v2} that there is a strong correlation between the event plane angle and that of the dijet momentum $\bm{q}_T$. Calculations of $v_n$ were then done using the direction of $\bm{q}_T$ as the event plane angle.

In spite of these overlapping aspects of ignorance, our modest range of  values for the Boer-Mulders' and Collins' functions seems to straightforwardly encapsulate the $v_2$ in $p$-$p$. Even more encouraging is the $A^{1/3}$ enhancement for $p$-$A$ collisions reproducing the larger $v_2$ in $p$-$Pb$. Our calculations show a mild tension with the measured $v_3$, and the comparisons with $v_4$ are inconclusive, as these have not been measured to a sufficiently high $p_T$.

We remind the reader that this is the first calculation of the azimuthal anisotropy at high $p_T$, arising from transverse momentum of the incoming partons in the initial state, in $p$-$p$ and $p$-$A$ collisions. As such, no solace should be drawn from the agreement with the measured $v_2$, and no alarm should be perceived in the tension with the measured $v_3$. The results of this paper, and the accompanying letter~\cite{Soudi:2023epi}, should be taken as an argument for further exploration of TMDPDFs as a source of azimuthal anisotropy in small system collisions. Without having done a full study, our na\"{i}ve expectation is that these effects will be diminished in systems with an extended final state, such as in heavy-ion collisions, where energy loss may completely wash out these initial state effects.  In fact, it may even be interesting to study the interference of this effect with the typical anisotropy from energy loss in semi-central collisions of intermediate size ions, e.g., $O$+$O$ collisions~\cite{Brewer:2021kiv}.

Restricting our attention to $p$-$p$ and $p$-$A$ collisions, we point out that all parton distribution functions used were leading twist and there could be sizable contributions at lower $p_T$ from higher twist contributions~\cite{Boer:2003cm}. These contributions could also be important for $v_3$ (and higher harmonics). As one moves towards a description of the remainder of the proton participating in the collision~\cite{JETSCAPE:2023xbc}, yet another contribution could originate in generalized transverse momentum dependent distributions~\cite{Boer:2018vdi}.

The $A^{1/3}$ enhancement in the case of $p$-$A$ is only known for the case of polarization averaged scattering, and has to be carried out for polarization dependent scattering. Also,  one should carefully check for the possibility of energy loss in the case of multiple scattering of partons in a large nucleus prior to the hard scattering that leads to high $p_T$ jets~\cite{Fries:2000da,Guo:1997it}. These calculations will rigorously test the hypothesised broadening without energy loss, required for the existence of an elliptic anisotropy with no modification to the angle integrated spectrum. 

Most prior calculations of multiple scattering of partons in nuclear matter~\cite{Fries:2000da,Guo:1997it,Wang:2001ifa,Fries:2002mu,Majumder:2007hx} have not considered gluons as the hard parton. For the currently measured $p_T$ range, gluon scattering is the most dominant effect, and thus, many of these calculations will have to be repeated for gluons. While gluon-gluon scattering is by far the largest contribution in the considered $p_T$ range, quark initiated processes will begin to become comparable at higher $p_T$ (see the Appendix). Quark mass effects (not necessarily for heavy-flavors) may allow for additional enhancement~\cite{Pisano:2013cya}. 

With further improvement in semi-analytical calculations, these effects will eventually have to be studied via extensive simulations, where both energy-momentum and angular momentum correlations between the hard and soft sectors can be thoroughly investigated. 
Any simulation, similar to the semi-analytic calculations in this paper, will have to assume factorization between hard and soft sectors in transverse momentum dependent high-$p_T$ hadron production. Such studies will of course suffer from the lack of a formal proof of factorization in traverse momentum dependent processes in $p$-$p$ (and by extension $p$-$A$) collisions~\cite{Rogers:2010dm}.

\section{Acknowledgments:}
 A. M. is indebted to S.~Pratt and J.~Putschke for discussions and for suggesting hard scattering as a source of the elliptic anisotropy. 
 Both authors are members of the JETSCAPE collaboration and have benefited from internal discussions. This work is supported in part by the U.S. Department of Energy under Grant No.~{DE-SC0013460}.
 I.~S. was funded as a part of the European Research Council project ERC-2018-ADG-835105 YoctoLHC, and as a part of the Center of Excellence in Quark Matter of the Academy of Finland (project 346325).

\appendix
\begin{figure}
	\begin{center}
		\includegraphics[width=0.45\textwidth]{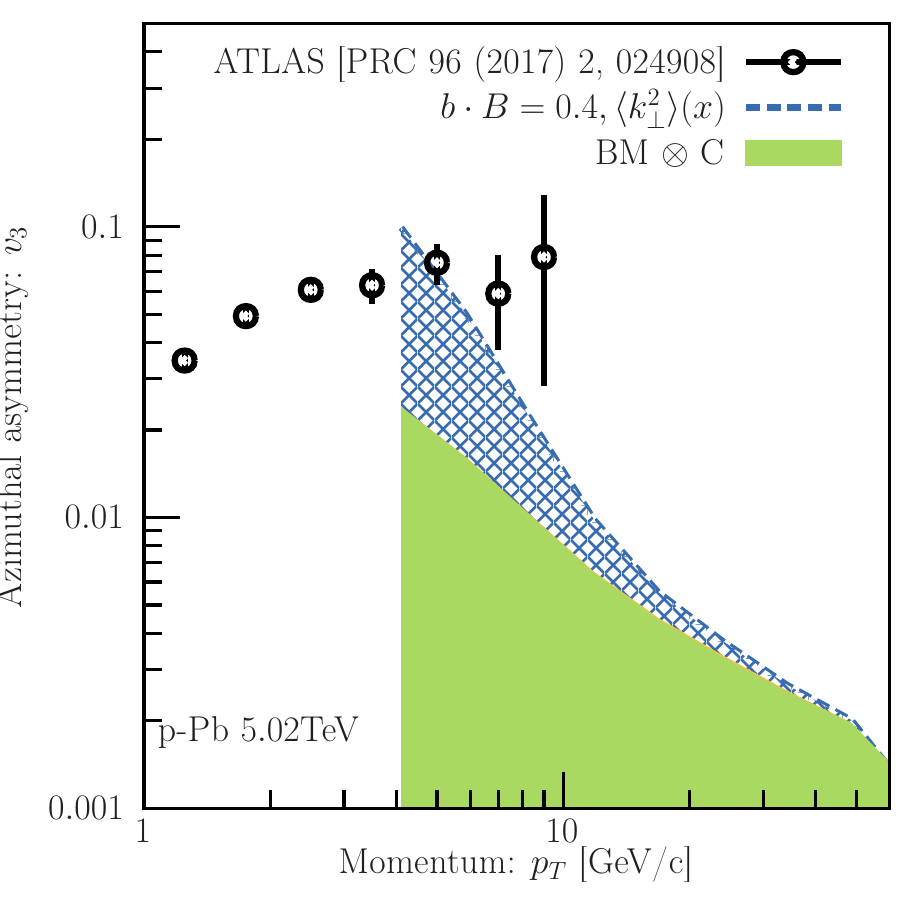}
		\includegraphics[width=0.45\textwidth]{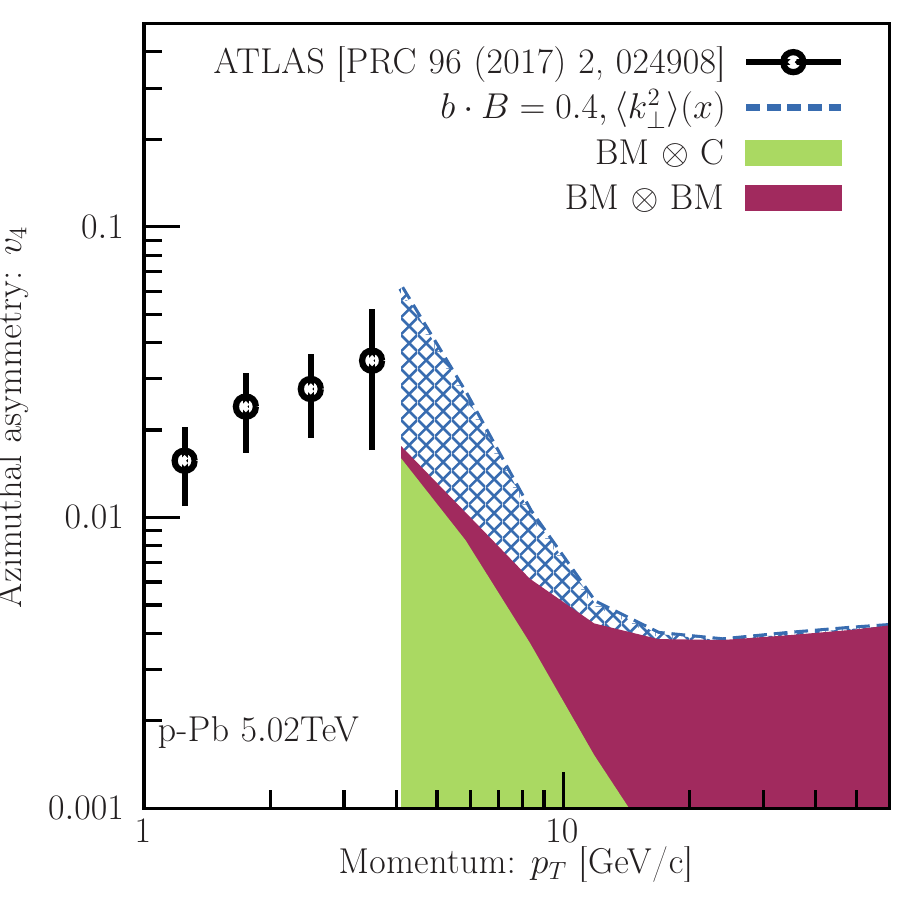}
	\end{center}
	\caption{
		Same as Fig.~\ref{fig:v2Log} but for $v_3$ (top) and $v_4$ (bottom).
		The red filled area is for $BM \otimes BM$ contribution.
	}
	\label{fig:v34pPb}
\end{figure}
\section{Quark Scattering}\label{app:QuarkScattering}
\begin{figure}
	\includegraphics[width=0.45\textwidth]{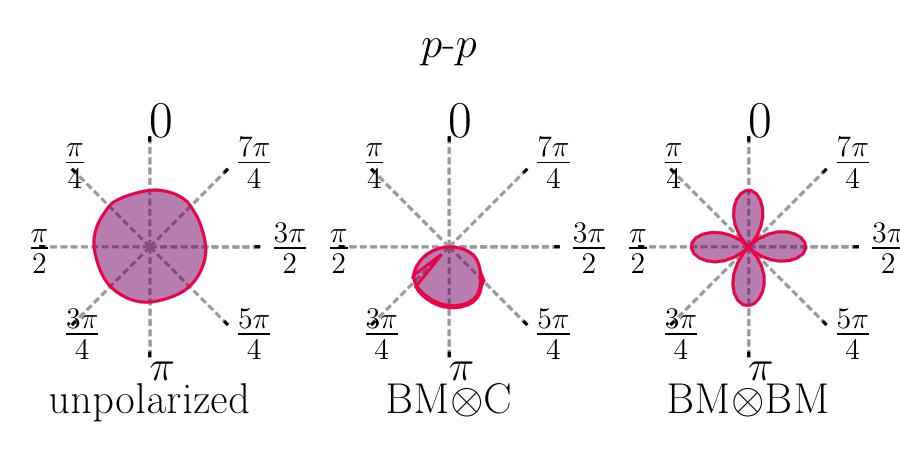}
	\includegraphics[width=0.45\textwidth]{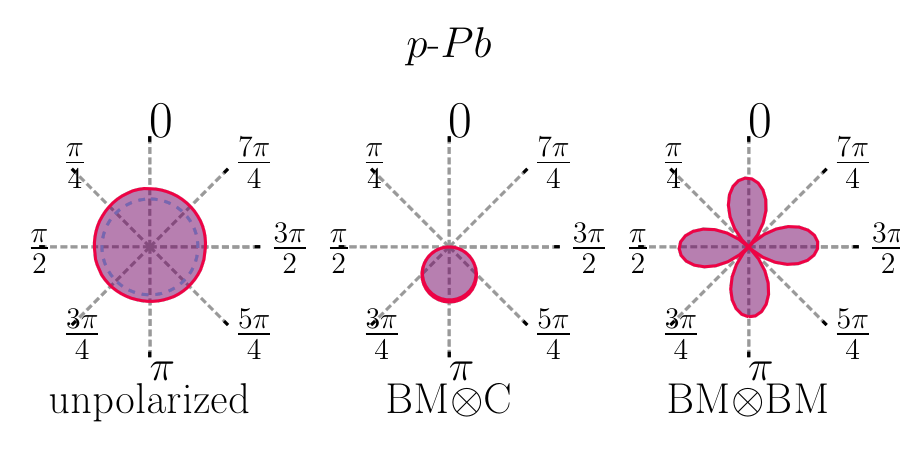}
	\caption{
		Azimuthal angle distribution $\frac{d\sigma}{d\phi}$ for $qq\to qq$ scattering pion production in $p$-$p$ at $5.02$ TeV (top) and $p$-$Pb$ at $8.16$ TeV (bottom) at a fixed $p_T=100$~GeV.
		The panels from left to right represents the unpolarized, $BM \otimes C$ and $BM \otimes BM$ contributions.
	}\label{fig:PhiDistqq}
\end{figure}
\begin{figure}
	\includegraphics[width=0.45\textwidth]{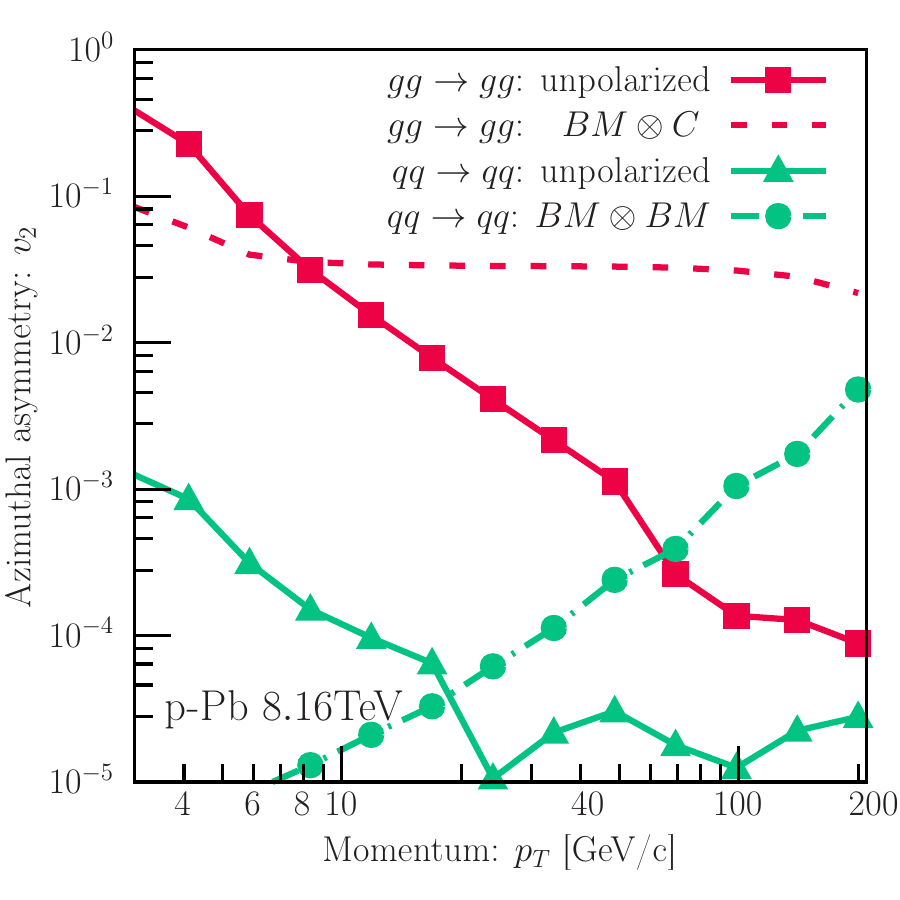}
	\caption{
		Azimuthal anisotropy $v_2$ for pion production in $p$-$Pb$ at $8.16$ TeV using a fixed $\langle k^2 \rangle^{1/2} = 1$ GeV and $B\cdot b = 0.4$ for the unpolarized distribution.
		The blue lines are the $gg\to gg$ contribution, the green lines are the $qq\to qq$ contribution.
		Full lines represent the unpolarized contribution, dashed lines represent $BM\otimes C$ terms and dashed lines circles represent $BM\otimes BM$ terms.
	}\label{fig:v2Process}
\end{figure}

Since the gluon PDFs and FFs are dominant at small scales, we have found that gluon-gluon scattering is the only relevant partonic scattering at $p_T$ range of the experimental results.
For completeness, we present the results obtained for quark scattering at high-$p_T$.
In Fig.~\ref{fig:PhiDistqq}, we present the azimuthal angle distribution $\frac{d\sigma}{d\phi}$ for $qq\to qq$ scattering in $p$-$p$ and $p$-$Pb$ collisions.
While the unpolarized contribution displays a similar shape as the gluon scattering, the polarized contributions are dominated by lower order coefficients than the gluon scattering.
For the quark the $BM\otimes C$ terms are dominated by $v_1$, while the $BM\otimes BM$ terms are dominated by $v_2$.

Computing the azimuthal anisotropy $v_2$, we find that the $qq\to qq$ scattering is subdominant compared to the gluon scattering as shown in Fig.~\ref{fig:v2Process}.
However, the $BM \otimes BM$ term is increasing with $p_T$ only surpassing the gluon contribution at $p_T \gg 200$~GeV.

\section{Kinematics of the cross section integration}
While we use a different approach to obtain the matrix element, the kinematics we use are based on \cite{DAlesio:2004eso,Contogouris:1978kh,Anselmino:2005sh}.
We reproduce the kinematics here.

We use the hadronic center of mass reference frame with Mandelstamn variables $(s,t,u)$, where the hadronic momenta are given by
\begin{align}
	p_A^+ 
	=& -p_B^-
	=
	\sqrt{\frac{s}{2}}\;,\quad
	p_A^- = p_B^+ = 0\;,\\
	p_C 
	=& \left(\sqrt{p_T^2 + p_L^2}, p_T, 0, p_L\right)\;.
\end{align}

The partonic scattering $a + b \to c + d$ is 
\begin{align}
	p_a =&
	p_A + \left(\frac{k_{\perp a}^2}{2x_a \sqrt{s}}, k_{\perp a}\cos\phi_a, k_{\perp a}\sin\phi_a, -\frac{k_{\perp a}^2}{2x_a \sqrt{s}}\right)\;,\\
	p_b =&
	p_B + \left(\frac{k_{\perp b}^2}{2x_b \sqrt{s}}, k_{\perp b}\cos\phi_b, k_{\perp b}\sin\phi_b, \frac{k_{\perp b}^2}{2x_b \sqrt{s}}\right)\;,
\end{align}

The detected hadron $C$ has light-cone momentum fraction $z=\frac{p_C^+}{p_c^+}$ and transverse momentum $\bm{k}_{\perp C}$ with respect to the parton $c$.
Using the delta function $\delta(\bm{k}_{\bot C}\cdot \hat{p}_c)$ from Eq.~(\ref{eq:cross-section}), we find that the partonic momentum is given by 
\begin{align}
	E_c
	=& \frac{E_C + \sqrt{E_C^2 - k_{\bot C}^2}}{2z}\;,\\
	\bm{p}_c
	=& \frac{E_c}{\sqrt{E_C^2-k^2_{\bot C}}}
	(\bm{p}_C - \bm{k}_{\bot C})\;.
\end{align}
The integration phase space can be written as
\begin{align}
	&\int d\Omega
	=\int \frac{dx_a dx_b dz}{2 \pi^2 z^3 s} d^2k_{\bot a} d^2 k_{\bot b} d^3 k_{\bot C}  \nonumber\\
	&\delta(\bm{k}_{\bot C}\cdot \hat{p}_c)J(\bm{k}_{\bot C})
	\delta(\hat{s} + \hat{t} + \hat{u})\;.
\end{align}
We define arbitrary transverse momentum $\bm{k}_{\bot C}$ in the hadronic center of mass frame as follows
\begin{align}
	&\bm{k}_{\bot C} \\
	&= k_{\bot C}\left(\sin\theta_{k_{\bot C}} \cos\phi_{k_{\bot C}},\sin\theta_{k_{\bot C}}\sin\phi_{k_{\bot C}},\cos\theta_{k_{\bot C}} \right)\;.\nonumber
\end{align}
The momentum of the outgoing parton $c$ is related to the pion's momentum $\bm{p}_\pi = p_T(1,0,\sinh y)$ as follows
\begin{align}
	\bm{p}_c 
	=& \bm{p}_\pi - \bm{k}_{\bot C}\;.
\end{align}
The transverse momentum delta function can be performed to define the azimuthal angle of $\bm{k}_{\bot C}$
\begin{align}
	\cos\phi_{k_{\bot C}}
	=&  \frac{k_{\bot C} - p_T \sinh y \cos \theta_{k_{\bot C}}}{pT \sin\theta_{k_{\bot C}}}\;,\\
	\sin \phi_{k_{\bot C}}^\pm
	=& \pm \sqrt{1-\cos^2\phi_{k_{\bot C}}}\;,
\end{align}
leading to the Jacobian 
\begin{align}
	&\int d^3 \bm{k}_{\perp C} \delta(\bm{k}_{\bot C}\cdot \hat{p}_c) f(\bm{k}_{\bot C}) \nonumber\\
	&= \int k_C dk_C d\theta_{k_C} 
	\frac{\sqrt{p_C^2 - k_C^2}}{p_T|\sin \phi_{k_C}|} 
	\left[f(\bm{k}_{\bot C}^{+}) + f(\bm{k}_{\bot C}^{-})\right]\;,
\end{align}
where one needs to sum over the two solutions $\sin \phi_{k_C}^\pm$.

Using the definition of the hadronic momentum, we obtain the Jacobian $J(\bm{k}_{\bot C})$ for the transformation from the parton $c$ to the hadron $C$ as follows
\begin{align}
	\frac{d^3 p_c}{E_c}
	= \frac{1}{z^2} J(\bm{k}_{\bot C}) \frac{d^3 p_C}{E_C}\;,
\end{align}
obtaining
\begin{align}
	J(\bm{k}_{\bot C}) = \frac{(E_C^2+\sqrt{\bm{p}_C^2-\bm{k}_{\bot C}^2})^2}{4(\bm{p}_C^2-\bm{k}_{\bot C}^2)}\;.
\end{align}

\bibliographystyle{physrev5}
\bibliography{ref}

\begin{thebibliography}{100}

\bibitem{STAR:2004jwm}
J.~Adams {\em et~al.}, STAR,
\newblock Phys. Rev. C {\bf 72}, 014904 (2005), arXiv:nucl-ex/0409033.

\bibitem{PHENIX:2003qra}
S.~S. Adler {\em et~al.}, PHENIX,
\newblock Phys. Rev. Lett. {\bf 91}, 182301 (2003), arXiv:nucl-ex/0305013.

\bibitem{ALICE:2010suc}
K.~Aamodt {\em et~al.}, ALICE,
\newblock Phys. Rev. Lett. {\bf 105}, 252302 (2010), arXiv:1011.3914.

\bibitem{CMS:2012zex}
S.~Chatrchyan {\em et~al.}, CMS,
\newblock Phys. Rev. C {\bf 87}, 014902 (2013), arXiv:1204.1409.

\bibitem{ATLAS:2012at}
G.~Aad {\em et~al.}, ATLAS,
\newblock Phys. Rev. C {\bf 86}, 014907 (2012), arXiv:1203.3087.

\bibitem{Kolb:2000sd}
P.~F. Kolb, J.~Sollfrank, and U.~W. Heinz,
\newblock Phys. Rev. {\bf C62}, 054909 (2000), hep-ph/0006129.

\bibitem{Wang:2003mm}
X.-N. Wang,
\newblock Phys. Lett. B {\bf 595}, 165 (2004), arXiv:nucl-th/0305010.

\bibitem{Majumder:2006we}
A.~Majumder,
\newblock Phys. Rev. {\bf C75}, 021901 (2007), arXiv:nucl-th/0608043.

\bibitem{Bass:2008rv}
S.~A. Bass {\em et~al.},
\newblock Phys.Rev. {\bf C79}, 024901 (2009), arXiv:0808.0908.

\bibitem{ATLAS:2018ezv}
M.~Aaboud {\em et~al.}, ATLAS,
\newblock Eur. Phys. J. C {\bf 78}, 997 (2018), arXiv:1808.03951.

\bibitem{ATLAS:2013ssy}
G.~Aad {\em et~al.}, ATLAS,
\newblock Phys. Rev. Lett. {\bf 111}, 152301 (2013), arXiv:1306.6469.

\bibitem{ALICE:2015efi}
J.~Adam {\em et~al.}, ALICE,
\newblock Phys. Lett. B {\bf 753}, 511 (2016), arXiv:1509.07334.

\bibitem{ATLAS:2012tjt}
G.~Aad {\em et~al.}, ATLAS,
\newblock Phys. Lett. B {\bf 719}, 220 (2013), arXiv:1208.1967.

\bibitem{CMS:2011iwn}
S.~Chatrchyan {\em et~al.}, CMS,
\newblock Phys. Rev. C {\bf 84}, 024906 (2011), arXiv:1102.1957.

\bibitem{ALICE:2013dpt}
B.~Abelev {\em et~al.}, ALICE,
\newblock JHEP {\bf 03}, 013 (2014), arXiv:1311.0633.

\bibitem{ATLAS:2015qmb}
G.~Aad {\em et~al.}, ATLAS,
\newblock JHEP {\bf 09}, 050 (2015), arXiv:1504.04337.

\bibitem{CMS:2012aa}
S.~Chatrchyan {\em et~al.}, CMS,
\newblock Eur. Phys. J. C {\bf 72}, 1945 (2012), arXiv:1202.2554.

\bibitem{ALICE:2010yje}
K.~Aamodt {\em et~al.}, ALICE,
\newblock Phys. Lett. B {\bf 696}, 30 (2011), arXiv:1012.1004.

\bibitem{STAR:2003fka}
J.~Adams {\em et~al.}, STAR,
\newblock Phys. Rev. Lett. {\bf 91}, 172302 (2003), arXiv:nucl-ex/0305015.

\bibitem{PHENIX:2001hpc}
K.~Adcox {\em et~al.}, PHENIX,
\newblock Phys. Rev. Lett. {\bf 88}, 022301 (2002), arXiv:nucl-ex/0109003.

\bibitem{Burke:2013yra}
K.~M. Burke {\em et~al.}, JET Collaboration,
\newblock Phys. Rev. {\bf C90}, 014909 (2014), arXiv:1312.5003.

\bibitem{Majumder:2010qh}
A.~Majumder and M.~Van~Leeuwen,
\newblock Prog. Part. Nucl. Phys. {\bf 66}, 41 (2011), arXiv:1002.2206.

\bibitem{Cao:2024pxc}
S.~Cao, A.~Majumder, R.~Modarresi-Yazdi, I.~Soudi, and Y.~Tachibana,
\newblock Int. J. Mod. Phys. E {\bf 33}, 2430002 (2024), arXiv:2401.10026.

\bibitem{Noronha-Hostler:2016eow}
J.~Noronha-Hostler, B.~Betz, J.~Noronha, and M.~Gyulassy,
\newblock Phys. Rev. Lett. {\bf 116}, 252301 (2016), arXiv:1602.03788.

\bibitem{Kumar:2019uvu}
A.~Kumar, A.~Majumder, and C.~Shen,
\newblock Phys. Rev. C {\bf 101}, 034908 (2020), arXiv:1909.03178.

\bibitem{He:2022evt}
Y.~He {\em et~al.},
\newblock Phys. Rev. C {\bf 106}, 044904 (2022), arXiv:2201.08408.

\bibitem{Baier:2002tc}
R.~Baier,
\newblock Nucl. Phys. {\bf A715}, 209 (2003), arXiv:hep-ph/0209038.

\bibitem{Majumder:2012sh}
A.~Majumder,
\newblock Phys. Rev. {\bf C87}, 034905 (2013).

\bibitem{Kumar:2020wvb}
A.~Kumar, A.~Majumder, and J.~H. Weber,
\newblock Phys. Rev. D {\bf 106}, 034505 (2022), arXiv:2010.14463.

\bibitem{Majumder:2008zg}
A.~Majumder,
\newblock Phys. Rev. {\bf C80}, 031902 (2009), arXiv:0810.4967.

\bibitem{Song:2010mg}
H.~Song, S.~A. Bass, U.~Heinz, T.~Hirano, and C.~Shen,
\newblock Phys. Rev. Lett. {\bf 106}, 192301 (2011), arXiv:1011.2783,
\newblock [Erratum: Phys. Rev. Lett.109,139904(2012)].

\bibitem{Shen:2014vra}
C.~Shen {\em et~al.},
\newblock Comput. Phys. Commun. {\bf 199}, 61 (2016), arXiv:1409.8164.

\bibitem{Bernhard:2019bmu}
J.~E. Bernhard, J.~S. Moreland, and S.~A. Bass,
\newblock Nature Phys. {\bf 15}, 1113 (2019).

\bibitem{Moreland:2018gsh}
J.~S. Moreland, J.~E. Bernhard, and S.~A. Bass,
\newblock Phys. Rev. C {\bf 101}, 024911 (2020), arXiv:1808.02106.

\bibitem{ATLAS:2017hap}
M.~Aaboud {\em et~al.}, ATLAS,
\newblock Eur. Phys. J. C {\bf 77}, 428 (2017), arXiv:1705.04176.

\bibitem{ALICE:2014dwt}
B.~B. Abelev {\em et~al.}, ALICE,
\newblock Phys. Rev. C {\bf 90}, 054901 (2014), arXiv:1406.2474.

\bibitem{CMS:2015yux}
V.~Khachatryan {\em et~al.}, CMS,
\newblock Phys. Rev. Lett. {\bf 115}, 012301 (2015), arXiv:1502.05382.

\bibitem{Dumitru:2010iy}
A.~Dumitru {\em et~al.},
\newblock Phys. Lett. B {\bf 697}, 21 (2011), arXiv:1009.5295.

\bibitem{Kovner:2012jm}
A.~Kovner and M.~Lublinsky,
\newblock Int. J. Mod. Phys. E {\bf 22}, 1330001 (2013), arXiv:1211.1928.

\bibitem{Gyulassy:2014cfa}
M.~Gyulassy, P.~Levai, I.~Vitev, and T.~S. Biro,
\newblock Phys. Rev. D {\bf 90}, 054025 (2014), arXiv:1405.7825.

\bibitem{Lappi:2015vta}
T.~Lappi, B.~Schenke, S.~Schlichting, and R.~Venugopalan,
\newblock JHEP {\bf 01}, 061 (2016), arXiv:1509.03499.

\bibitem{Altinoluk:2015uaa}
T.~Altinoluk, N.~Armesto, G.~Beuf, A.~Kovner, and M.~Lublinsky,
\newblock Phys. Lett. B {\bf 751}, 448 (2015), arXiv:1503.07126.

\bibitem{Dumitru:2015gaa}
A.~Dumitru, T.~Lappi, and V.~Skokov,
\newblock Phys. Rev. Lett. {\bf 115}, 252301 (2015), arXiv:1508.04438.

\bibitem{Altinoluk:2016vax}
T.~Altinoluk, N.~Armesto, G.~Beuf, A.~Kovner, and M.~Lublinsky,
\newblock Phys. Rev. D {\bf 95}, 034025 (2017), arXiv:1610.03020.

\bibitem{Hagiwara:2017ofm}
Y.~Hagiwara, Y.~Hatta, B.-W. Xiao, and F.~Yuan,
\newblock Phys. Lett. B {\bf 771}, 374 (2017), arXiv:1701.04254.

\bibitem{Altinoluk:2018ogz}
T.~Altinoluk, N.~Armesto, A.~Kovner, and M.~Lublinsky,
\newblock Eur. Phys. J. C {\bf 78}, 702 (2018), arXiv:1805.07739.

\bibitem{Altinoluk:2020psk}
T.~Altinoluk, N.~Armesto, A.~Kovner, M.~Lublinsky, and V.~V. Skokov,
\newblock Eur. Phys. J. C {\bf 81}, 583 (2021), arXiv:2012.01810.

\bibitem{Hatta:2020bgy}
Y.~Hatta, B.-W. Xiao, F.~Yuan, and J.~Zhou,
\newblock Phys. Rev. Lett. {\bf 126}, 142001 (2021), arXiv:2010.10774.

\bibitem{ATLAS:2016yzd}
M.~Aaboud {\em et~al.}, ATLAS,
\newblock Phys. Rev. C {\bf 96}, 024908 (2017), arXiv:1609.06213.

\bibitem{ATLAS:2019vcm}
G.~Aad {\em et~al.}, ATLAS,
\newblock Eur. Phys. J. C {\bf 80}, 73 (2020), arXiv:1910.13978.

\bibitem{CMS:2016xef}
V.~Khachatryan {\em et~al.}, CMS,
\newblock JHEP {\bf 04}, 039 (2017), arXiv:1611.01664.

\bibitem{ALICE:2015umm}
J.~Adam {\em et~al.}, ALICE,
\newblock Phys. Lett. B {\bf 749}, 68 (2015), arXiv:1503.00681.

\bibitem{ATLAS:2022kqu}
ATLAS,
\newblock (2022), arXiv:2211.15257.

\bibitem{Collins:1981uw}
J.~C. Collins and D.~E. Soper,
\newblock Nucl. Phys. {\bf B194}, 445 (1982).

\bibitem{Boer:1997nt}
D.~Boer and P.~J. Mulders,
\newblock Phys. Rev. D {\bf 57}, 5780 (1998), arXiv:hep-ph/9711485.

\bibitem{Collins:1992kk}
J.~C. Collins,
\newblock Nucl. Phys. B {\bf 396}, 161 (1993), arXiv:hep-ph/9208213.

\bibitem{Bierlich:2022pfr}
C.~Bierlich {\em et~al.},
\newblock (2022), arXiv:2203.11601.

\bibitem{Collins:1985ue}
J.~C. Collins, D.~E. Soper, and G.~Sterman,
\newblock Nucl. Phys. {\bf B261}, 104 (1985).

\bibitem{Boer:2010zf}
D.~Boer, S.~J. Brodsky, P.~J. Mulders, and C.~Pisano,
\newblock Phys. Rev. Lett. {\bf 106}, 132001 (2011), arXiv:1011.4225.

\bibitem{Pisano:2013cya}
C.~Pisano, D.~Boer, S.~J. Brodsky, M.~G.~A. Buffing, and P.~J. Mulders,
\newblock JHEP {\bf 10}, 024 (2013), arXiv:1307.3417.

\bibitem{denDunnen:2014kjo}
W.~J. den Dunnen, J.~P. Lansberg, C.~Pisano, and M.~Schlegel,
\newblock Phys. Rev. Lett. {\bf 112}, 212001 (2014), arXiv:1401.7611.

\bibitem{Boer:2014tka}
D.~Boer and W.~J. den Dunnen,
\newblock Nucl. Phys. B {\bf 886}, 421 (2014), arXiv:1404.6753.

\bibitem{Bacchetta:1999kz}
A.~Bacchetta, M.~Boglione, A.~Henneman, and P.~J. Mulders,
\newblock Phys. Rev. Lett. {\bf 85}, 712 (2000), arXiv:hep-ph/9912490.

\bibitem{Anselmino:2004ky}
M.~Anselmino, M.~Boglione, U.~D'Alesio, E.~Leader, and F.~Murgia,
\newblock Phys. Rev. D {\bf 71}, 014002 (2005), arXiv:hep-ph/0408356.

\bibitem{Anselmino:2005sh}
M.~Anselmino {\em et~al.},
\newblock Phys. Rev. D {\bf 73}, 014020 (2006), arXiv:hep-ph/0509035.

\bibitem{DAlesio:2004eso}
U.~D'Alesio and F.~Murgia,
\newblock Phys. Rev. D {\bf 70}, 074009 (2004), arXiv:hep-ph/0408092.

\bibitem{Mulders:2000sh}
P.~J. Mulders and J.~Rodrigues,
\newblock Phys. Rev. D {\bf 63}, 094021 (2001), arXiv:hep-ph/0009343.

\bibitem{Field:1989uq}
R.~D. Field,
\newblock Redwood City, USA: Addison-Wesley (1989) 366 p. (Frontiers in physics, 77).

\bibitem{BermudezMartinez:2018fsv}
A.~Bermudez~Martinez {\em et~al.},
\newblock Phys. Rev. D {\bf 99}, 074008 (2019), arXiv:1804.11152.

\bibitem{BermudezMartinez:2021lxz}
A.~Bermudez~Martinez, F.~Hautmann, and M.~L. Mangano,
\newblock Phys. Lett. B {\bf 822}, 136700 (2021), arXiv:2107.01224.

\bibitem{Arnold:1983mw}
R.~G. Arnold {\em et~al.},
\newblock Phys. Rev. Lett. {\bf 52}, 727 (1984).

\bibitem{Cronin:1974zm}
J.~W. Cronin {\em et~al.}, E100,
\newblock Phys. Rev. D {\bf 11}, 3105 (1975).

\bibitem{Gamberg:2005ip}
L.~P. Gamberg and G.~R. Goldstein,
\newblock Phys. Lett. B {\bf 650}, 362 (2007), arXiv:hep-ph/0506127.

\bibitem{Guo:1997it}
X.~Guo,
\newblock Phys. Rev. D {\bf 58}, 036001 (1998), arXiv:hep-ph/9711453.

\bibitem{Fries:2002mu}
R.~J. Fries,
\newblock Phys. Rev. D {\bf 68}, 074013 (2003), arXiv:hep-ph/0209275.

\bibitem{Majumder:2007hx}
A.~Majumder and B.~Muller,
\newblock Phys. Rev. {\bf C77}, 054903 (2008), arXiv:0705.1147.

\bibitem{Kovchegov:1999yj}
Y.~V. Kovchegov,
\newblock Phys. Rev. D {\bf 60}, 034008 (1999), arXiv:hep-ph/9901281.

\bibitem{Mueller:1989st}
A.~H. Mueller,
\newblock Nucl. Phys. B {\bf 335}, 115 (1990).

\bibitem{Gribov:1983ivg}
L.~V. Gribov, E.~M. Levin, and M.~G. Ryskin,
\newblock Phys. Rept. {\bf 100}, 1 (1983).

\bibitem{Arnold:2008vd}
P.~Arnold and W.~Xiao,
\newblock Phys. Rev. {\bf D78}, 125008 (2008), arXiv:0810.1026.

\bibitem{Soudi:2023epi}
I.~Soudi and A.~Majumder,
\newblock Phys. Lett. B {\bf 859}, 139105 (2024), arXiv:2308.14702.

\bibitem{Collins_2011}
J.~Collins,
\newblock {\em Foundations of Perturbative QCD}$\mbox{}$\,\,Cambridge Monographs on Particle Physics, Nuclear Physics and Cosmology (Cambridge University Press, 2011).

\bibitem{Kniehl:2000fe}
B.~A. Kniehl, G.~Kramer, and B.~Potter,
\newblock Nucl. Phys. {\bf B582}, 514 (2000), hep-ph/0010289.

\bibitem{Aybat:2011zv}
S.~M. Aybat and T.~C. Rogers,
\newblock Phys. Rev. D {\bf 83}, 114042 (2011), arXiv:1101.5057.

\bibitem{Barry:2023qqh}
P.~C. Barry {\em et~al.}, Jefferson Lab Angular Momentum (JAM),
\newblock Phys. Rev. D {\bf 108}, L091504 (2023), arXiv:2302.01192.

\bibitem{Soffer:1994ww}
J.~Soffer,
\newblock Phys. Rev. Lett. {\bf 74}, 1292 (1995), arXiv:hep-ph/9409254.

\bibitem{Mantysaari:2017cni}
H.~M\"antysaari, B.~Schenke, C.~Shen, and P.~Tribedy,
\newblock Phys. Lett. B {\bf 772}, 681 (2017), arXiv:1705.03177.

\bibitem{Buttar:2005gdq}
C.~M. Buttar {\em et~al.},
\newblock {The Underlying Event},
\newblock in {\em {HERA and the LHC: A Workshop on the Implications of HERA for LHC Physics: CERN - DESY Workshop 2004/2005 (Midterm Meeting, CERN, 11-13 October 2004; Final Meeting, DESY, 17-21 January 2005)}}, pp. 192--217, Geneva, 2005, CERN.

\bibitem{Sjostrand:1987su}
T.~Sjostrand and M.~van Zijl,
\newblock Phys. Rev. D {\bf 36}, 2019 (1987).

\bibitem{Andersson:1983ia}
B.~Andersson, G.~Gustafson, G.~Ingelman, and T.~Sjostrand,
\newblock Phys. Rept. {\bf 97}, 31 (1983).

\bibitem{Borghini:2001vi}
N.~Borghini, P.~M. Dinh, and J.-Y. Ollitrault,
\newblock Phys. Rev. C {\bf 64}, 054901 (2001), arXiv:nucl-th/0105040.

\bibitem{ALICE:2019hno}
S.~Acharya {\em et~al.}, ALICE,
\newblock Phys. Rev. C {\bf 101}, 044907 (2020), arXiv:1910.07678.

\bibitem{ATLAS:2016xpn}
G.~Aad {\em et~al.}, ATLAS,
\newblock Phys. Lett. B {\bf 763}, 313 (2016), arXiv:1605.06436.

\bibitem{ALICE:2016dei}
J.~Adam {\em et~al.}, ALICE,
\newblock Phys. Lett. B {\bf 760}, 720 (2016), arXiv:1601.03658.

\bibitem{Kovarik:2015cma}
K.~Kovarik {\em et~al.},
\newblock Phys. Rev. D {\bf 93}, 085037 (2016), arXiv:1509.00792.

\bibitem{PHENIX:2021dod}
U.~A. Acharya {\em et~al.}, PHENIX,
\newblock Phys. Rev. C {\bf 105}, 064902 (2022), arXiv:2111.05756.

\bibitem{Brewer:2021kiv}
J.~Brewer, A.~Mazeliauskas, and W.~van~der Schee,
\newblock {Opportunities of OO and $p$O collisions at the LHC},
\newblock in {\em {Opportunities of OO and pO collisions at the LHC}}, 2021, arXiv:2103.01939.

\bibitem{Boer:2003cm}
D.~Boer, P.~J. Mulders, and F.~Pijlman,
\newblock Nucl. Phys. B {\bf 667}, 201 (2003), arXiv:hep-ph/0303034.

\bibitem{JETSCAPE:2023xbc}
Jetscape {\em et~al.}, JETSCAPE,
\newblock PoS {\bf HardProbes2023}, 128 (2024), arXiv:2308.02650.

\bibitem{Boer:2018vdi}
D.~Boer, T.~Van~Daal, P.~J. Mulders, and E.~Petreska,
\newblock JHEP {\bf 07}, 140 (2018), arXiv:1805.05219.

\bibitem{Fries:2000da}
R.~J. Fries, A.~Schafer, E.~Stein, and B.~Muller,
\newblock Nucl. Phys. B {\bf 582}, 537 (2000), arXiv:hep-ph/0002074.

\bibitem{Wang:2001ifa}
X.-N. Wang and X.-F. Guo,
\newblock Nucl. Phys. {\bf A696}, 788 (2001), arXiv:hep-ph/0102230.

\bibitem{Rogers:2010dm}
T.~C. Rogers and P.~J. Mulders,
\newblock Phys. Rev. D {\bf 81}, 094006 (2010), arXiv:1001.2977.

\bibitem{Contogouris:1978kh}
A.~P. Contogouris, R.~Gaskell, and S.~Papadopoulos,
\newblock Phys. Rev. D {\bf 17}, 2314 (1978).

\end{thebibliography}

\end{document}